\documentclass[aps,prl,groupedaddress,superscriptaddress,twocolumn,10pt,longbibliography]{revtex4-2}
\usepackage[utf8]{inputenc}
\usepackage{hyperref}
\usepackage{graphicx}
\graphicspath{{../figures/}}
\usepackage{setspace}
\usepackage{orcidlink}
\newcommand{\corauthor}[2]{
    \author{#1}
    \email{#2}
}

\usepackage{braket}
\usepackage{siunitx}
\usepackage{amsmath}
\begin{document}

\title{Simulating neural network criticality and resource dynamics with Rydberg gases}
\corauthor{Patrick Mischke\orcidlink{0000-0001-7859-8426}}{agott-publication@physik.rptu.de}
\affiliation{RPTU University Kaiserslautern-Landau, Department of Physics and State Research Center OPTIMAS, Kaiserslautern, Germany}
\affiliation{Max Planck Graduate Center with Johannes Gutenberg University Mainz (MPGC), 55128 Mainz, Germany}

\author{Herwig Ott\orcidlink{0000-0002-3155-2719}}
\affiliation{RPTU University Kaiserslautern-Landau, Department of Physics and State Research Center OPTIMAS, Kaiserslautern, Germany}

\author{Michael Fleischhauer\orcidlink{0000-0003-4059-7289}}
\affiliation{RPTU University Kaiserslautern-Landau, Department of Physics and State Research Center OPTIMAS, Kaiserslautern, Germany}

\author{Thomas Niederprüm\orcidlink{0000-0001-8336-4667}}
\affiliation{RPTU University Kaiserslautern-Landau, Department of Physics and State Research Center OPTIMAS, Kaiserslautern, Germany}

\date{\today}

\begin{abstract}
Efficient operation of neural networks has been linked to criticality in their underlying non-equilibrium excitation dynamics.
However, obtaining experimental evidence of this conjecture remains challenging due to limited control and undersampling in biological systems.
Here, we experimentally explore neural network criticality using an ultracold Rydberg gas as a highly controllable simulator.
We highlight the similarity of the excitation spreading via Rydberg facilitation and the synaptic connection of spiking activity of neurons, giving rise to distinct absorbing and active phases.
We systematically explore and resolve criticality criteria, including power-law scaling of excitation avalanches and the emergence of universal avalanche shape collapse.
Crucially, we implement a controlled gain mechanism to compensate for atom loss, mimicking metabolic resource replenishment and stabilizing the system in a controlled non-equilibrium steady state.
We find  peak temporal correlations at the critical point and stochastic oscillations with dragon king avalanches in the active phase, consistent with predictions for systems orbiting criticality.
Our work establishes facilitated Rydberg gases as a platform for investigating criticality, resource dynamics, and emergent oscillations in neural networks.
\end{abstract}

\maketitle


\section*{Introduction}

In recent years, converging evidence has been obtained that efficient operation of neural networks is connected to criticality in the underlying non-equilibrium excitation dynamics \cite{langtonComputationEdgeChaos1990, beggsNeuronalAvalanchesNeocortical2003, beggsCriticalityHypothesisHow2007, dearcangelisLearningPhenomenonOccurring2010, klausStatisticalAnalysesSupport2011, friedmanUniversalCriticalDynamics2012, shewFunctionalBenefitsCriticality2013, fagerholmCascadesCognitiveState2015}.
In particular, it has been shown that the ability of neural networks to react to very different input magnitudes (dynamic range) \cite{kinouchiOptimalDynamicalRange2006, shewNeuronalAvalanchesImply2009, gautamMaximizingSensoryDynamic2015}, the ability to detect very small input changes (susceptibility) \cite{ beggsCortexCriticalPoint2022}, and the ability to learn \cite{dearcangelisLearningPhenomenonOccurring2010} maximize in the presence of criticality.
Even more, the shortcomings of current artificial deep neural networks with respect to robustness, efficiency, and pathologies in comparison to their biological counterparts might indicate that operation in the vicinity of a critical point is possibly a crucial ingredient for proper function \cite{vockCriticalDynamicsGoverns2026}.
On the other hand, it has been questioned whether neural networks truly exhibit a sharp critical point combined with a self-organization mechanism, or whether an extended long-range ordered phase instead gives rise to a similar phenomenology \cite{sunMemoryNeuralActivity2025, siplingCriticalAssessmentBrain2026}.
Moreover, the dynamic adaptation of the network through consumption and replenishment of metabolic resources has been identified as a process relevant to the phase space dynamics in the vicinity of the critical point \cite{robertsCriticalRoleResource2014, virkarFeedbackControlStabilization2016, franovicCollectiveActivityBursting2022}.
It is thus a field of intense study to elucidate the role of criticality in such networks and how phenomena like, e.g., stochastic oscillations \cite{kinouchiStochasticOscillationsDragon2019} and dragon king avalanches \cite{dearcangelisAreDragonkingNeuronal2012, mishraDragonkinglikeExtremeEvents2018} arise in them.
While many of these phenomena are clearly visible in numerical models, it is challenging to experimentally study the role of criticality in physical realizations of neural networks.
At present, research on this important question is largely driven by comparing predictions of a broad set of numerical models with different initial assumptions to a rather small set of experimental data taken from biological samples.
However, these experiments typically suffer from undersampling and the lack of suitable fine-grained control parameters to judge how well a certain numerical model fits the studied system.

\begin{figure*}
    \includegraphics{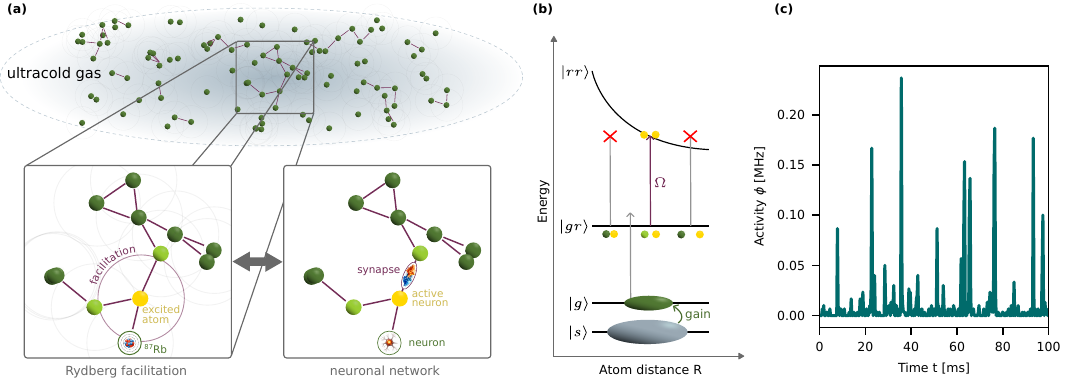}
    \caption{
        \textbf{(a)} The dynamics of a neural network is based on active neurons (yellow) that activate connected neurons (light green) via synaptic activation (purple), leading to cascaded, spiking activity of the network.
        Analogously, in an ultracold gas, an emergent local network (purple lines) is formed by those ground state atoms (green) that are approximately in the facilitation distance $r_\mathrm{fac}$ (purple circle).
        Through Rydberg facilitation an existing excitation (yellow) can efficiently spread to connected nodes (light green) of the network, leading to excitation cascades/avalanches in the network.
        \textbf{(b)} The synaptic process in the simulator arises through Rydberg facilitation which off-resonantly couples one atom from a pair of atoms in the $\ket{g}$ state to form the pair state $\ket{gr}$ or $\ket{rg}$.
        Due to the strong interaction between two Rydberg atoms, the energy of the doubly excited $\ket{rr}$ state depends on the distance between the atoms and the second excitation only occurs in the facilitation distance $r_\mathrm{fac}$ where the driving laser with Rabi frequency $\Omega$ becomes resonant.
        Via optical pumping atoms can be continuously transferred from the facilitation-inactive reservoir state $\ket{s}$ into the $\ket{g}$ state and replace lost atoms.
        \textbf{(c)} Typical integrated activity signal obtained from the simulator close to the critical point. The signal shows characteristic features such as non-Poissonian spiking patterns, burst-like avalanches and variable inter-spike intervals.
    }
    \label{fig:correspondence-sketch}
\end{figure*}

We address these challenges by building a physical simulator for critical neural network dynamics using a bottom-up approach.
By starting from well-controlled particles with tunable interactions we create an accessible and controllable many-body system that is able to simulate the essential features of neural network criticality and its dynamics.
To accurately implement the proper criticality features, our simulator exhibits the same kind of absorbing state phase transition, possesses an inhibition channel which scales with the activity, and has a controllable resource replenishment mechanism which stabilizes the operation near the critical point.
Here, we show that our simulator provides these ingredients through replicating the non-linear excitation spreading in neural networks in combination with a tailored gain process, giving rise to a non-equilibrium phase transition with high control over the microscopic degrees of freedom and a controlled resource replenishment which allows us to stabilize the system in the active phase and to study long-term dynamics of the system in the vicinity of the critical point.


\section*{Neural network simulator}

In our simulator, the most basic conceptual elements, the neurons, are replaced by the individual atoms in an ultracold gas (Fig.\,\ref{fig:correspondence-sketch}\,a).
The atoms can either be in the ground state $\ket{g}$, corresponding to an inactive neuron, or in the excited state $\ket{r}$, representing an active (firing) neuron.
The synaptic connection is realized by the Rydberg facilitation process \cite{atesAntiblockadeRydbergExcitation2007, amthorEvidenceAntiblockadeUltracold2010}, where an active atom, together with a suitable global driving laser, can induce the activation of another atom at the facilitation distance $r_\mathrm{fac} = \sqrt[6]{\frac{C_6}{\Delta}}$ (see Fig.\,\ref{fig:correspondence-sketch}\,b).
This facilitation criterion defines spherical shells around each active atom, in which the activation of other atoms can take place.
As a result, a network with dynamic and local connectivity emerges from those atoms in the gas that have a matching distance (Fig.\,\ref{fig:correspondence-sketch} a).
Since the atoms in our experiment are confined as an ultracold gas in a crossed optical dipole trap (see methods), the resulting network structure has the node degree distribution of an Erdős-Rényi random graph \cite{ruttenModelingRydbergGases2021, ohlerNonequilibriumUniversalityRydbergexcitation2025}.
Furthermore, the inherent dephasing in the facilitation-based offspring creation leads to a classical contact process \cite{leviQuantumNonequilibriumDynamics2016}.
The competition between facilitated excitation and decay of the Rydberg state maps the system onto the susceptible-infected-susceptible (SIS) epidemic model which features an absorbing state phase transition that shows a critical point \cite{ohlerNonequilibriumUniversalityRydbergexcitation2025, schemppFullCountingStatistics2014a, malossiFullCountingStatistics2014, marcuzziNonequilibriumUniversalityDynamics2015, helmrichSignaturesSelforganizedCriticality2020} and gives rise to activation avalanches and epidemic growth of excitations \cite{wintermantelEpidemicGrowthGriffiths2021}. Near the critical point of this phase transition, characteristic non-Poissonian spiking patterns and burst-like avalanches of activity (see Fig.\,\ref{fig:correspondence-sketch}\,c) emerge.

A biological neural network consumes metabolic resources that have to be replenished for stable long-term operation, e.g., via the glial network.
This process is known to play a decisive role in determining the system's critical behavior \cite{virkarFeedbackControlStabilization2016}.
It is therefore crucial to implement such a mechanism in a simulator.
As biological neural networks and facilitated Rydberg gases share strong conceptual similarities not only on a general level but also on a microscopic level (Fig.\,\ref{fig:correspondence-sketch}\,a), this can be achieved in a straightforward way: much like a neuron that can fire several times, a single atom can perform multiple cycles of excitation and decay to the ground state.
Eventually, an excited atom will, however, be ionized or decays to an uncoupled ground state $\ket{g_1}$ effectively removing it from the system.
Atom loss in our system serves as a functional proxy for the net reduction of excitable units in biological networks, which arises from refractory periods, synaptic depression, and metabolic constraints rather than cell loss.
As the atom density shrinks, the likelihood of excitation spreading decreases.
This loss mechanism leads to a self-organization process that pushes the system from the active phase towards the critical point \cite{helmrichSignaturesSelforganizedCriticality2020, klockeHydrodynamicStabilizationSelfOrganized2021} and, on a slower timescale, deeper into the absorbing state.
We compensate this loss process by adding a dedicated gain channel based on optical pumping, which stabilizes and fine-tunes the system in the vicinity of the critical point.
Both processes together allow us to capture critical neural network dynamics \cite{virkarDynamicRegulationResource2020, siplingCriticalAssessmentBrain2026} with a control knob for all relevant system parameters.

\begin{figure*}
    \includegraphics{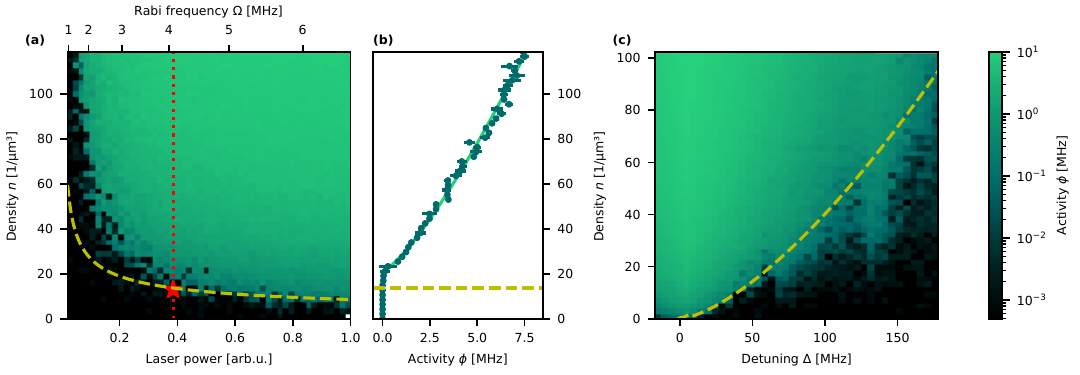}
    \caption{
        2D phase diagrams of the system for \textbf{(a)} varied particle density and driving laser power with fixed detuning $\Delta = 2\pi\cdot\SI{40}{\mega\hertz}$ or \textbf{(c)} driving laser detuning with fixed driving strength $\Omega = \SI{5.5}{\mega\hertz}$.
        The system is always prepared at a certain density $n$ (y-axis) and the measured activity $\phi$ for a \SI{1}{\milli\second} excitation pulse of Rabi frequency $\Omega$ and detuning $\Delta$ is shown  as color code.
        The red star indicates the position of the measurement shown in Fig.\,\ref{fig:meta-g2} with maximal fluctuations.
        In the central panel \textbf{(b)} a dedicated measurement along a vertical cut on a linear activity scale is shown along with a fit of the power-law scaling $\phi = (n-n_c)^\beta$ (light green), yielding $n_c = 20.9(7)\,\mathrm{\mu m}^{-3}$ and $\beta = 0.78(2)$.
        The yellow dashed line denotes the predicted phase transition in a parameter-free model of the spreading process (see Methods).
        Additional features arise in the $\Delta$-dependent measurement because pair states at $\Delta \approx \SI{135}{\mega\hertz}$ causing deviations from the $r^{-6}$ interaction.
    }
    \label{fig:phase-diagram}
    \label{fig:phase-diagram-cut}
\end{figure*}


\section*{Proximity to a Critical Point}
While well established frameworks exist to describe equilibrium phase transitions and the connected criticality, the non-equilibrium case has less stringent theoretical foundation.
In addition, the presence of dissipative processes like resource consumption, or atom loss in our case, complicates the situation as the system might not be able to actually reside at the critical point.
We here follow Ref. \cite{beggsCortexCriticalPoint2022}, which has put forward six criteria (I - VI) that have to be fulfilled in such systems.

The first three criteria impose the presence of distinct phases characterized by an order parameter (III), a control parameter to tune between them (II) and a transition between them at a branching ratio of 1 (I) where the branching ratio is the average number of offspring events per parent event in an avalanche and can thus serve as an indicator of whether a system is subcritical (ratio \textless{}~1), critical (ratio = 1), or supercritical (ratio \textgreater{}~1).
These criteria can be simultaneously demonstrated by carefully measuring the phase diagram and comparing it to the calculated branching ratio (see methods).

The experimental phase diagrams (see methods for an explanation of the experimental protocol) are obtained by varying different control parameters and measuring the system activity $\phi$ (Fig.\,\ref{fig:phase-diagram}).
One can clearly identify the phase transition between the absorbing phase, where the activity is close to zero, and the active phase, where it is large.
Thereby, the activity serves as the order parameter of the phase transition (criterion (III)).

To show that the branching ratio at the transition is close to 1 (criterion (I)), we numerically calculate it in a mean-field model without any free fit parameters.
The details of the model are outlined in the methods section.
The transition point resulting from the model is shown as a yellow line in Fig.\,\ref{fig:phase-diagram} and is in good agreement with the measured onset of activity across a wide range of the phase diagram.

The ground state density $n$ takes the role of a control parameter that tunes across the critical point (criterion (II)).
The Rabi frequency $\Omega$ and the detuning $\Delta$ define the systems characteristics in a similar way, i.e. they tune the critical density $n_c$, and are experimentally easily accessible.
While tuning across the phase transition using these parameters is possible, additional effects such as power broadening and the vicinity of nearby Rydberg pair states make quantitative evaluations easier when choosing the  density as a control parameter.
Nevertheless, the ability to tune those additional two parameters in the Rydberg system enhances the control over the microscopic spreading process to a level which is hard or even impossible to achieve in other platforms.

\begin{figure*}
    \includegraphics{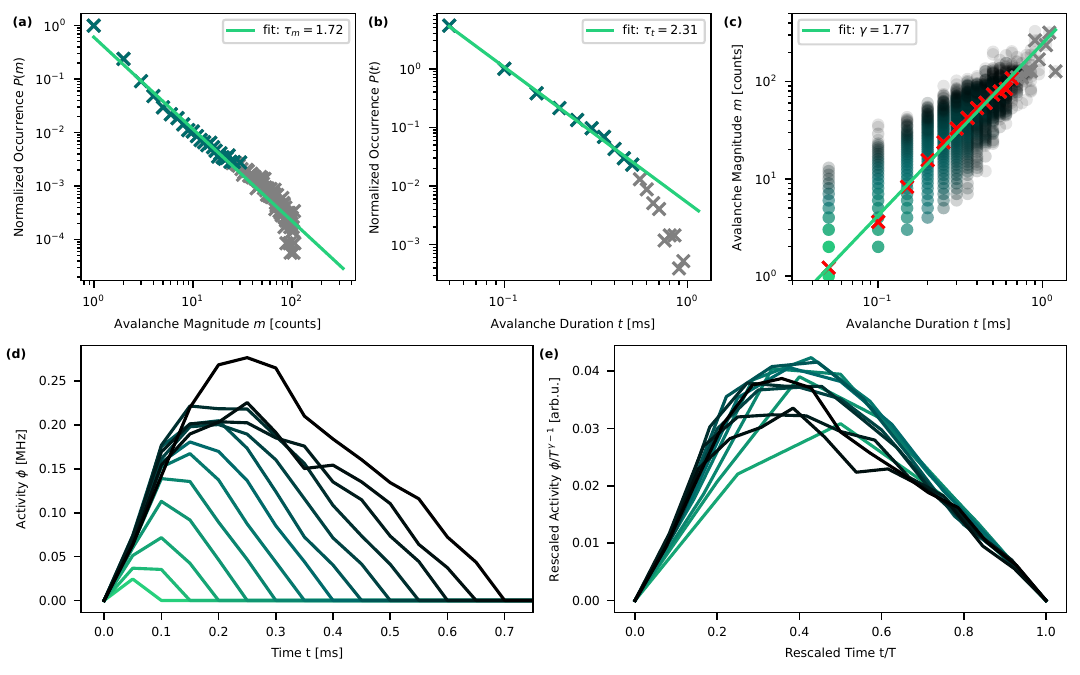}
    \caption{
        Top Row: Power-law behavior in the avalanche distributions close to the critical point:
        distribution of \textbf{(a)} avalanche magnitudes $P(m)$  and \textbf{(b)} avalanche durations $P(t)$.
        We fit power-law exponents of $\tau_m = 1.72$ and $\tau_t = 2.31$ (light green lines).
        Panel \textbf{(c)} shows all recorded avalanches with their respective duration $t$ and magnitude $m$ as green circles with the crosses denoting the mean value $\langle m\rangle_t$ for each discrete duration $t$.
        We fit a power-law exponent of $\gamma = 1.77$ (light green line).
        Data points with insufficient statistics (gray) have been omitted for fitting the exponent.
        Bottom row: Avalanche shapes close to the critical point.
        \textbf{(d)} Averaged signal of all avalanches of a specific duration $T$ ranging $T=\SI{0.1}{\milli\second}$ (bright green) to $T=\SI{0.7}{\milli\second}$.
        \textbf{(e)} Collapse of the avalanche shapes shown in (d) by normalizing the duration and rescaling the amplitude by $\phi/T^{\gamma -1}$ with the value for $\gamma$ obtained in (c).
        Due to missing temporal resolution, the first two durations have been omitted.}
    \label{fig:avalanche-power-laws}
    \label{fig:avalanche-shapes}
\end{figure*}

Three more criteria target the scale-invariance of a system at criticality by requiring the existence of power laws (IV), a defined scaling relation between them (V), and a universal scaling function to collapse, e.g., avalanche shapes (VI).
Observing power-laws is often considered one of the key features of criticality \cite{helmrichSignaturesSelforganizedCriticality2020}.
In addition to the scaling of the order parameter, $\phi = (n-n_c)^\beta$ (see fit in Fig.\,\ref{fig:phase-diagram-cut}), excitation avalanches are expected to show power-law scaling:
The distributions $P(m)$ and $P(t)$ of avalanches with magnitude $m$ or duration $t$ are scaling with the exponents $P(m)\sim m^{-\tau_m}$ and $P(t)\sim t^{-\tau_t}$.
The relation between magnitude and duration of an avalanche should also follow a power-law with $\langle m\rangle_t \sim t^\gamma$, with $\gamma \approx \frac{\tau_t-1}{\tau_m-1}$ \cite{sethnaCracklingNoise2001, markovicPowerLawsSelforganized2014}.
Fig.\,\ref{fig:avalanche-power-laws}\,(a) and (b) show the avalanche distributions when preparing the system at a density close to the critical density.
Both distributions follow power laws and fulfill the exponent relation for  $\gamma$ (criterion (IV) and (V)).
The measured $\beta$ exponent is compatible with the expectations for (anomalous) directed percolation within the error bounds \cite{bradyAnomalousDirectedPercolation2024}, while the avalanche exponents are above theoretically predicted values.
Universal exponents in non-equilibrium transitions can depend on additional parameters.
For example, we have shown previously that the avalanche exponents increase from the expected value when dissipation is present \cite{ohlerNonequilibriumUniversalityRydbergexcitation2025}.
In this regard our experiment qualitatively agrees well with observations on biological neural networks that frequently show varying avalanche exponents, while the exponent relations remains valid \cite{fosqueEvidenceQuasicriticalBrain2021}.
Such systems were named quasi-critical, as they are near a critical point, but don't allow for direct extraction of universal exponents.

Scale invariance close to the critical point implies that certain self-similar patterns are observable when looking at different sizes of the same observable.
Most prominently, the temporal activity pattern of an avalanche, averaged over all avalanches with a certain duration $T$, should be independent of its duration.
Our experimental results are displayed in Fig.\,\ref{fig:avalanche-shapes}\,(e), where we scale all measured avalanches to the same duration and show that the activity, rescaled with $\frac{1}{T^{\gamma-1}}$, collapses on a single curve (criterion (VI)).
The avalanche shape resembles a skewed parabola that has been observed in other systems as well \cite{spasojevicBarkhausenNoiseElementary1996, sethnaCracklingNoise2001, fagerholmCascadesCognitiveState2015,friedmanUniversalCriticalDynamics2012}.
These results demonstrate that facilitated Rydberg gases possess all the necessary properties to simulate critical dynamics in realistic neural networks.

\section*{Resource Replenishment}

\begin{figure*}
    \includegraphics{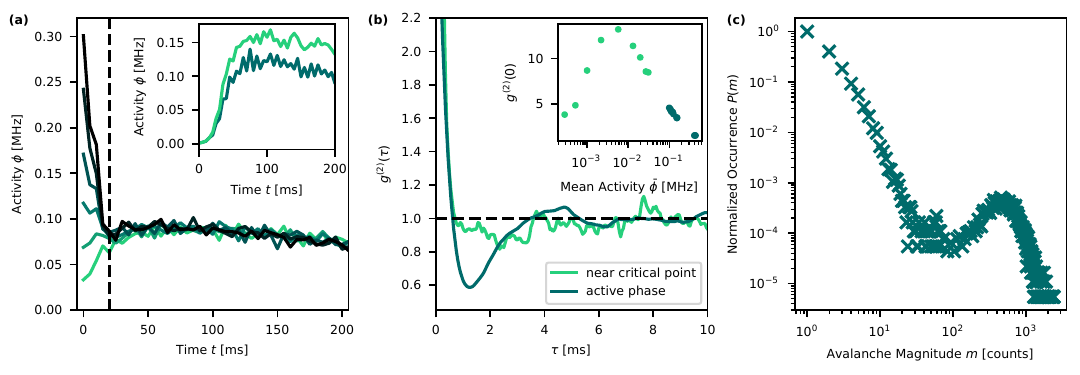}
    \caption{
        \textbf{(a)}
        Long-term evolution of the system's activity averaged across 90 realizations per curve with a constant gain rate.
        The collapse of the average activity for different starting densities after $t\approx\SI{20}{\milli\second}$ (dashed black line) shows that the system is attracted towards and stabilizes at the dynamical steady state where the applied constant gain equals the intrinsic loss.
        The inset shows the opposing case: Different steady states can be achieved with different applied gain.
        \textbf{(b)} Temporal correlation function $g^{(2)}(\tau)$ extracted at steady states close to criticality (light green, taken at position of the marker in Fig.\,\ref{fig:phase-diagram}\,a) and slightly in the active phase (dark green).
        Close to criticality the correlations show strong bunching of $g^{(2)}(\tau=0) = 13.2$ with a characteristic timescale much longer than the lifetime of a single Rydberg excitation.
        Additionally, in the active phase, these correlations are followed by a clear anti-bunching for delays between $\tau\approx \SI{0.5}{\milli\second}$ and $\tau\approx \SI{4}{\milli\second}$, indicating stochastic oscillations in the activity of the system.
        Inset: $g^{(2)}(\tau=0)$ at different points in the phase diagram.
        Light green points without gain, dark green points with gain stabilization.
        The correlations peak close to the critical point.
        Since the density that we reach at the steady state is not always known, we characterize the position across the phase transition by the mean activity $\bar{\phi}$ of the steady state.
        \textbf{(c)}
        Dragon king avalanches appearing in the magnitude distribution in the active phase.
        Avalanches with $m \approx 500$ occur orders of magnitude more frequently than expected from an extrapolated power law.
        To improve visibility, points with $m > 67$ show averages for intervals of width 10.}
    \label{fig:stabilization}
    \label{fig:g2}
    \label{fig:meta-g2}
    \label{fig:dragon-kings}
\end{figure*}

To accurately simulate the long-term criticality dynamics of biological neural networks around the critical point it is necessary to also consider resource replenishment processes that counteract dissipation.
Multiple mechanisms like inhibitory neurons provide short term stability and protection against run-away activity, but metabolic resource distribution is a key factor on a timescale much longer than the firing dynamics of the neurons \cite{virkarFeedbackControlStabilization2016}.
Analogous to this metabolic resource consumption, atom loss in our Rydberg system takes place on a longer timescale than the facilitation dynamics and reduces the average probability to spread an excitation.
We compensate for this resource consumption, i.e. atom loss, by using the facilitation-inactive $F=1$ ground state as a reservoir from which we optically pump atoms into the $F=2$ state that can be facilitated (see Fig.\,\ref{fig:correspondence-sketch}\,b).
Movement of the atoms compensates local density variations and has been described as hydrodynamic stabilization mechanism for short time scales \cite{klockeHydrodynamicStabilizationSelfOrganized2021}, while the pump process controls the global density on longer time scales.
Adjusting the effective gain rate via the strength of optical pumping, we tune the equilibrium between consumption and replenishment of resources and thus set the position of the steady state in the phase diagram.
For a given gain, the system then moves towards the same steady state irrespective of the initial position set by differing initial densities (see Fig.\,\ref{fig:stabilization}\,a).
Similarly, it evolves from a common starting point to different steady states if different effective gain rates are set (see inset of Fig.\,\ref{fig:stabilization}\,a).

With this crucial ingredient implemented as an additional tuning knob, we study the system's long-term dynamics by analyzing the temporal correlations of the driven system in the steady state.
Fig.\,\ref{fig:g2} shows the $g^{(2)}(\tau)$ temporal correlation function of firing events for two different points in the phase diagram: one close to the critical point and one slightly in the active phase.
For both points a strong instantaneous correlation of excitation events (bunching) at $\tau=0$ appears, reflecting the excitation cascades driven by the non-linear synaptic connection implemented through facilitation.
Analyzing this bunching behavior in $g^{(2)}(\tau=0)$ across the phase transition reveals that the correlations as a function of the mean activity $\bar{\phi}$ show a clear peak structure (inset in Fig.\,\ref{fig:meta-g2}).
The position of maximum fluctuations is shown as red marker in the phase diagram Fig.\,\ref{fig:phase-diagram}.
It is both close to the critical point predicted by the branching ratio calculations and the steep increase in the activity from \SI{e-3}{\mega\hertz} to \SI{e-1}{\mega\hertz}.
This demonstrates that the amount of correlations is a suitable measure to pinpoint the phase transition in neural networks.

While the strength of the correlations maximizes at the critical point,
correlation times do not show such a pronounced peak structure due to limited separation of time scales between facilitation, decay and pump processes.
In the vicinity of the critical point, this correlation relaxes on a timescale of $\tau \approx \SI{150}{\micro\second}$, much longer than the lifetime of an individual Rydberg atom ($\approx\SI{50}{\micro\second}$), but compatible with the average loss timescale after multiple excitation cycles.
In the active phase, the correlation drops on a similar timescale but does not directly relax to an uncorrelated state.
It rather shows an oscillatory behavior that turns into an anti-correlation for an even longer timescale of $\tau_\mathrm{repl} \approx \SI{2}{\milli\second}$.
As seen from Fig.\,\ref{fig:stabilization}, the averaged activity across multiple experiment repetitions is constant, indicating that these oscillations are a stochastic and intrinsic property of the system and not externally induced.
This oscillatory behavior around the steady state driven by stochastic perturbations was numerically predicted to occur in neural networks when the system orbits the critical point \cite{kinouchiStochasticOscillationsDragon2019, friesRhythmsCognitionCommunication2015,zangStructuralConstraintsEmergence2024, marenduzzoOscillationSIRSModel2025, aliakbarianTransitionSelforganizedCriticality2026}.
These arise when, following a stochastic seed event in an over-critical system, the activity increases rapidly and the associated dissipation causes the density to rapidly fall below the critical point - the avalanche leaves a depleted region behind.
When the outburst eventually ends, the system is below the critical point and starts to accumulate density on a longer timescale due to the gain mechanism, eventually enabling a start of the next oscillatory cycle triggered by a stochastically appearing seed event.
As known from the numerical simulations \cite{kinouchiStochasticOscillationsDragon2019}, this behavior is associated with so-called dragon king avalanches: Outbursts with a large magnitude, that occur more frequently than an extrapolation of the probability distribution of the smaller events would predict.
When analyzing the probability distribution for different avalanche magnitudes slightly in the active phase (Fig.\,\ref{fig:dragon-kings}\,c), we can demonstrate also this phenomenon predicted to appear in neuronal networks to be present in our simulator.


\section*{Discussion}

Our findings establish facilitated Rydberg gases as a versatile experimental platform for dissecting the interplay between criticality, resource dynamics, and emergent long-term features in neural networks.
The ability to map out the phase diagram helps to understand how self-organizing dynamics allow a physical system to operate near the critical point.
Our work opens a way of qualitatively investigating stochastic oscillations and the appearance of dragon king avalanches.
Rapid progress on tweezer platforms for ultracold atoms \cite{kaufmanQuantumScienceOpticalTweezer2021} opens the prospect of creating networks with full control over the network topology and the read-out.
While the measurements presented here rely on classical excitation spreading due to strong dephasing, the platform inherently supports coherent dynamics.
Future implementations tailored for coherent interactions could thus enable the study of quantum neural networks with $10^5$ nodes, regimes that remain out of reach for classical simulation.
In particular, this would allow studying the interplay between criticality and quantum entanglement.
Quantum entanglement in neural networks has been suggested as a relevant property in biological settings \cite{kerskensExperimentalIndicationsNonclassical2022, liuEntangledBiphotonGeneration2024, ExploringConsciousnessPhoton2025}, but remains an open question.
Furthermore, the quantum mechanical nature of the facilitation process could be harnessed to improve computational features of the network \cite{bravoQuantumReservoirComputing2022}.
Ultimately, this approach offers a pathway to significantly deepen our understanding of the fundamental working principles of neural networks and non-equilibrium critical systems.

\section*{Methods}
\subsection*{Experimental Apparatus}
To take the presented measurements we employ two-stage laser cooling in a 2D-MOT and 3D-MOT to prepare a cigar-shaped sample of ${}^{87}\mathrm{Rb}$ atoms inside a crossed optical dipole trap where we drive forced evaporation until we end up with $N=5\times 10^5$ atoms at trapping frequencies of $\omega_{x,y,z} = 2\pi\times(425, 115, 439)\,\si{\hertz}$ and a temperature of $T=\SI{2.25}{\micro\kelvin}$.
All atoms are initially in the reservoir state $\ket{g_0} = \ket{s}\equiv\ket{5\mathrm{S}_{1/2}, \mathrm{F}=1}$.

Using optical pumping via the $5\mathrm{P}_{3/2}$ state, we can transfer atoms in the $\ket{g}\equiv \ket{5\mathrm{S}_{1/2}, \mathrm{F}=2}$ hyperfine ground state.
The repumping laser from the initial MOT phase serves as optical pumping laser.
Due to the better control over this parameter, the detuning of the pumping laser is used to control the pumping rate and a calibration measurement allows for a direct correspondence between laser detuning and global pumping rate $\gamma$.
When keeping up a constant gain rate over extended periods of time, the reservoir state $\ket{s}$ is depleted and thus the absolute pumping rate drops.
To compensate for this, the pumping laser is ramped according to $\gamma(t) = \frac{\gamma_0}{1-\gamma_0 t}$ to obtain a constant gain rate $\gamma_0$ for an extended time.

The $\ket{g}$ state is coupled to the $\ket{r}\equiv 40\mathrm{P}_{3/2}$ Rydberg state via a single photon transition by a $\SI{297}{\nano\meter}$ excitation laser that is focused to a Gaussian waist of $w_0 = \SI{100}{\micro\meter}$ and linearly polarized.
External magnetic fields are compensated during the measurement such that the linear polarization of the excitation laser sets the quantization axis.
The laser globally illuminates the sample at an angle of 45° with respect to the long axis of the sample.
To allow for Rydberg facilitation and to suppress resonant excitation, a blue detuning of $\Delta=2\pi\cdot(-15\dots +\SI{175}{\mega\hertz})$ to the atomic Rydberg resonance can be set.
All measurements are taken with $\Delta=2\pi\cdot\SI{40}{\mega\hertz}$ unless noted otherwise.
This leads to an adjustable facilitation distance of
$r_\mathrm{fac} \approx \SI{1.2}{\micro\meter}$ (for $40\mathrm{P}_{3/2}$ and $\Delta = \SI{40}{\mega\hertz}$, with $C_6 = \SI{0.111}{\giga\hertz\micro\meter^6}$ calculated using the pairinteraction software package \cite{weberCalculationRydbergInteraction2017}).
By changing the coupling laser power, the resonant coupling strength for atoms in the facilitation distance can be set in the range $\Omega=0\dots\SI{6.5}{\mega\hertz}$.
The coupling strength $\Omega$ is calibrated by performing microwave spectroscopy on the $\ket{s}\rightarrow\ket{g}$ transition and observing the light shift for different coupling laser intensities.

While the detuning strongly suppresses excitations of atoms without a facilitation partner, off-resonant events  still occur occasionally.
In comparison to neural networks, where input is typically from an external source, this results in an intrinsic seed process for excitations in the Rydberg system.

The loss of Rydberg atoms into ions through photoionization and associative ionization not only mimics the resource consumption, it also serves as a detection channel for the activity of the network, picking up discrete events comparable to electrodes for observing neural networks.
Rydberg atoms that eventually decay to ions after several iterations of excitation and decay, are guided to a time-resolved dynode detector with detection efficiencies of $\eta_\mathrm{det}\approx40\%$ using a small electric bias field.
The primary measurement signal is generated by binning the arrival of ions into bins with \SI{50}{\micro\second} duration.
The activity in counts per time unit is directly proportional to the Rydberg density $n_\mathrm{Ryd}$, where the proportionality factor $\eta < 1$ depends on the decay rate, detection efficiency and trapping volume.
Our time-resolved measurement scheme with single count sensitivity allows us to identify individual avalanches, which is required to obtain the avalanche exponents $\tau$ and $\alpha$.
To discriminate between avalanches, we consider an avalanche to end whenever there is an empty bin.

To measure the phase diagrams, we use the optical pumping process to prepare our system with ground state density $n$ in the coupled state $\ket{g}$ while the excitation laser is off and subsequently switch on the excitation pulse at a set Rabi frequency $\Omega$ and detuning $\Delta$ for a fixed duration of \SI{2}{\milli\second}.
This duration is chosen longer than the initial growth dynamics \cite{wintermantelEpidemicGrowthGriffiths2021} and short enough that the density change due to atom loss, i.e. resource consumption, is limited.
Using absorption imaging in a time-of-flight scheme, we can measure atom number and temperature of the prepared sample in an independent measurement without excitation pulses.

\subsection*{Calculation of Branching-Ratio}
We consider the case of a test atom initially in the ground state and at distance $r$ from the seed atom, which decays with rate $\gamma$.
The dynamics of the test atom can be described in terms of the imbalance $w=\rho_{gg} - \rho_{ee}$ as given by the optical Bloch equations \cite{nohAnalyticSolutionsOptical2010}, where
$w(t) = w(t, \Omega, \gamma, \gamma^*, \Delta')$ depends on the Rabi frequency $\Omega$, local detuning $\Delta'(r)$, decay rate $\gamma$ and dephasing $\gamma^*$.
This allows to obtain the probability $P_\mathrm{fac}$ that the test atom is in the excited state at the time of decay:
$$P_\mathrm{fac} = \int_{0}^\infty \frac{1-w}{2}\gamma e^{-\gamma t}dt$$
To obtain the branching ratio $BR$, we calculate the number of atoms a given seed atom will have excited via the facilitation mechanism at the time of decay.
In a mean-field picture without correlations, this number can be obtained by integrating the probability $P_\mathrm{fac}$ over the volume of the system and the density $n$ of ground state atoms.
The local detuning is given by the laser detuning $\Delta$ and the Rydberg-Rydberg interaction as $\Delta'(r) = \Delta - \frac{C_6}{r^6}$:
$$BR = \int_{R^3} n \left[P_\mathrm{fac}\left(\Omega, \gamma, \gamma*, \Delta'(r)\right) - P_\mathrm{fac}\left(\Omega, \gamma, \gamma*, \Delta\right)\right]dV$$
As we are only interested in excitations via the facilitation mechanism, we subtract any off-resonant excitations that occur independently of the distance to our seed atom.
We numerically solve this equation for $BR = 1$ to obtain the critical density $n_c$.
A more detailed derivation is given in the supplementary material.

\section*{Acknowledgements}
We would like to thank Volker Scheuss for helpful discussions and proofreading the manuscript.
The authors acknowledge financial support by the DFG within the collaborative research center TR 185 OSCAR (Number 277625399).
This work was also supported by the research initiative Quantum Computing for Artificial Intelligence (QC-AI) and by the Max Planck Graduate Center MPGC with the
University of Mainz.

\section*{Data Availability}
The data that support the plots within this paper and other findings of this study are publicly available \cite{data}.

\section*{Author Contributions}
P.M. performed the experiments and analyzed the data. P.M. and T.N. prepared the manuscript.
T.N. and H.O. and M.F. developed the concepts for the study.
T.N. and H.O. conceived and supervised the experiment.
All authors contributed to the data interpretation and final manuscript preparation.

\section*{Competing financial interests}
The authors declare no competing financial interests.

%


\begin{thebibliography}{50}%
\makeatletter
\providecommand \@ifxundefined [1]{%
 \@ifx{#1\undefined}
}%
\providecommand \@ifnum [1]{%
 \ifnum #1\expandafter \@firstoftwo
 \else \expandafter \@secondoftwo
 \fi
}%
\providecommand \@ifx [1]{%
 \ifx #1\expandafter \@firstoftwo
 \else \expandafter \@secondoftwo
 \fi
}%
\providecommand \natexlab [1]{#1}%
\providecommand \enquote  [1]{``#1''}%
\providecommand \bibnamefont  [1]{#1}%
\providecommand \bibfnamefont [1]{#1}%
\providecommand \citenamefont [1]{#1}%
\providecommand \href@noop [0]{\@secondoftwo}%
\providecommand \href [0]{\begingroup \@sanitize@url \@href}%
\providecommand \@href[1]{\@@startlink{#1}\@@href}%
\providecommand \@@href[1]{\endgroup#1\@@endlink}%
\providecommand \@sanitize@url [0]{\catcode `\\12\catcode `\$12\catcode
  `\&12\catcode `\#12\catcode `\^12\catcode `\_12\catcode `\%12\relax}%
\providecommand \@@startlink[1]{}%
\providecommand \@@endlink[0]{}%
\providecommand \url  [0]{\begingroup\@sanitize@url \@url }%
\providecommand \@url [1]{\endgroup\@href {#1}{\urlprefix }}%
\providecommand \urlprefix  [0]{URL }%
\providecommand \Eprint [0]{\href }%
\providecommand \doibase [0]{https://doi.org/}%
\providecommand \selectlanguage [0]{\@gobble}%
\providecommand \bibinfo  [0]{\@secondoftwo}%
\providecommand \bibfield  [0]{\@secondoftwo}%
\providecommand \translation [1]{[#1]}%
\providecommand \BibitemOpen [0]{}%
\providecommand \bibitemStop [0]{}%
\providecommand \bibitemNoStop [0]{.\EOS\space}%
\providecommand \EOS [0]{\spacefactor3000\relax}%
\providecommand \BibitemShut  [1]{\csname bibitem#1\endcsname}%
\let\auto@bib@innerbib\@empty
\bibitem [{\citenamefont {Langton}(1990)}]{langtonComputationEdgeChaos1990}%
  \BibitemOpen
  \bibfield  {author} {\bibinfo {author} {\bibfnamefont {C.~G.}\ \bibnamefont
  {Langton}},\ }\bibfield  {title} {\bibinfo {title} {Computation at the edge
  of chaos: {{Phase}} transitions and emergent computation},\ }\href
  {https://doi.org/10.1016/0167-2789(90)90064-V} {\bibfield  {journal}
  {\bibinfo  {journal} {Physica D: Nonlinear Phenomena}\ }\textbf {\bibinfo
  {volume} {42}},\ \bibinfo {pages} {12} (\bibinfo {year} {1990})}\BibitemShut
  {NoStop}%
\bibitem [{\citenamefont {Beggs}\ and\ \citenamefont
  {Plenz}(2003)}]{beggsNeuronalAvalanchesNeocortical2003}%
  \BibitemOpen
  \bibfield  {author} {\bibinfo {author} {\bibfnamefont {J.~M.}\ \bibnamefont
  {Beggs}}\ and\ \bibinfo {author} {\bibfnamefont {D.}~\bibnamefont {Plenz}},\
  }\bibfield  {title} {\bibinfo {title} {Neuronal {{Avalanches}} in
  {{Neocortical Circuits}}},\ }\href
  {https://doi.org/10.1523/JNEUROSCI.23-35-11167.2003} {\bibfield  {journal}
  {\bibinfo  {journal} {Journal of Neuroscience}\ }\textbf {\bibinfo {volume}
  {23}},\ \bibinfo {pages} {11167} (\bibinfo {year} {2003})}\BibitemShut
  {NoStop}%
\bibitem [{\citenamefont {Beggs}(2007)}]{beggsCriticalityHypothesisHow2007}%
  \BibitemOpen
  \bibfield  {author} {\bibinfo {author} {\bibfnamefont {J.~M.}\ \bibnamefont
  {Beggs}},\ }\bibfield  {title} {\bibinfo {title} {The criticality hypothesis:
  How local cortical networks might optimize information processing},\ }\href
  {https://doi.org/10.1098/rsta.2007.2092} {\bibfield  {journal} {\bibinfo
  {journal} {Philosophical Transactions of the Royal Society A: Mathematical,
  Physical and Engineering Sciences}\ }\textbf {\bibinfo {volume} {366}},\
  \bibinfo {pages} {329} (\bibinfo {year} {2007})}\BibitemShut {NoStop}%
\bibitem [{\citenamefont {{de Arcangelis}}\ and\ \citenamefont
  {Herrmann}(2010)}]{dearcangelisLearningPhenomenonOccurring2010}%
  \BibitemOpen
  \bibfield  {author} {\bibinfo {author} {\bibfnamefont {L.}~\bibnamefont {{de
  Arcangelis}}}\ and\ \bibinfo {author} {\bibfnamefont {H.~J.}\ \bibnamefont
  {Herrmann}},\ }\bibfield  {title} {\bibinfo {title} {Learning as a phenomenon
  occurring in a critical state},\ }\href
  {https://doi.org/10.1073/pnas.0912289107} {\bibfield  {journal} {\bibinfo
  {journal} {Proceedings of the National Academy of Sciences}\ }\textbf
  {\bibinfo {volume} {107}},\ \bibinfo {pages} {3977} (\bibinfo {year}
  {2010})}\BibitemShut {NoStop}%
\bibitem [{\citenamefont {Klaus}\ \emph {et~al.}(2011)\citenamefont {Klaus},
  \citenamefont {Yu},\ and\ \citenamefont
  {Plenz}}]{klausStatisticalAnalysesSupport2011}%
  \BibitemOpen
  \bibfield  {author} {\bibinfo {author} {\bibfnamefont {A.}~\bibnamefont
  {Klaus}}, \bibinfo {author} {\bibfnamefont {S.}~\bibnamefont {Yu}},\ and\
  \bibinfo {author} {\bibfnamefont {D.}~\bibnamefont {Plenz}},\ }\bibfield
  {title} {\bibinfo {title} {Statistical {{Analyses Support Power Law
  Distributions Found}} in {{Neuronal Avalanches}}},\ }\href
  {https://doi.org/10.1371/journal.pone.0019779} {\bibfield  {journal}
  {\bibinfo  {journal} {PLOS ONE}\ }\textbf {\bibinfo {volume} {6}},\ \bibinfo
  {pages} {e19779} (\bibinfo {year} {2011})}\BibitemShut {NoStop}%
\bibitem [{\citenamefont {Friedman}\ \emph {et~al.}(2012)\citenamefont
  {Friedman}, \citenamefont {Ito}, \citenamefont {Brinkman}, \citenamefont
  {Shimono}, \citenamefont {DeVille}, \citenamefont {Dahmen}, \citenamefont
  {Beggs},\ and\ \citenamefont
  {Butler}}]{friedmanUniversalCriticalDynamics2012}%
  \BibitemOpen
  \bibfield  {author} {\bibinfo {author} {\bibfnamefont {N.}~\bibnamefont
  {Friedman}}, \bibinfo {author} {\bibfnamefont {S.}~\bibnamefont {Ito}},
  \bibinfo {author} {\bibfnamefont {B.~A.~W.}\ \bibnamefont {Brinkman}},
  \bibinfo {author} {\bibfnamefont {M.}~\bibnamefont {Shimono}}, \bibinfo
  {author} {\bibfnamefont {R.~E.~L.}\ \bibnamefont {DeVille}}, \bibinfo
  {author} {\bibfnamefont {K.~A.}\ \bibnamefont {Dahmen}}, \bibinfo {author}
  {\bibfnamefont {J.~M.}\ \bibnamefont {Beggs}},\ and\ \bibinfo {author}
  {\bibfnamefont {T.~C.}\ \bibnamefont {Butler}},\ }\bibfield  {title}
  {\bibinfo {title} {Universal {{Critical Dynamics}} in {{High Resolution
  Neuronal Avalanche Data}}},\ }\href
  {https://doi.org/10.1103/PhysRevLett.108.208102} {\bibfield  {journal}
  {\bibinfo  {journal} {Physical Review Letters}\ }\textbf {\bibinfo {volume}
  {108}},\ \bibinfo {pages} {208102} (\bibinfo {year} {2012})}\BibitemShut
  {NoStop}%
\bibitem [{\citenamefont {Shew}\ and\ \citenamefont
  {Plenz}(2013)}]{shewFunctionalBenefitsCriticality2013}%
  \BibitemOpen
  \bibfield  {author} {\bibinfo {author} {\bibfnamefont {W.~L.}\ \bibnamefont
  {Shew}}\ and\ \bibinfo {author} {\bibfnamefont {D.}~\bibnamefont {Plenz}},\
  }\bibfield  {title} {\bibinfo {title} {The {{Functional Benefits}} of
  {{Criticality}} in the {{Cortex}}},\ }\href
  {https://doi.org/10.1177/1073858412445487} {\bibfield  {journal} {\bibinfo
  {journal} {The Neuroscientist}\ }\textbf {\bibinfo {volume} {19}},\ \bibinfo
  {pages} {88} (\bibinfo {year} {2013})}\BibitemShut {NoStop}%
\bibitem [{\citenamefont {Fagerholm}\ \emph {et~al.}(2015)\citenamefont
  {Fagerholm}, \citenamefont {Lorenz}, \citenamefont {Scott}, \citenamefont
  {Dinov}, \citenamefont {Hellyer}, \citenamefont {Mirzaei}, \citenamefont
  {Leeson}, \citenamefont {Carmichael}, \citenamefont {Sharp}, \citenamefont
  {Shew},\ and\ \citenamefont {Leech}}]{fagerholmCascadesCognitiveState2015}%
  \BibitemOpen
  \bibfield  {author} {\bibinfo {author} {\bibfnamefont {E.~D.}\ \bibnamefont
  {Fagerholm}}, \bibinfo {author} {\bibfnamefont {R.}~\bibnamefont {Lorenz}},
  \bibinfo {author} {\bibfnamefont {G.}~\bibnamefont {Scott}}, \bibinfo
  {author} {\bibfnamefont {M.}~\bibnamefont {Dinov}}, \bibinfo {author}
  {\bibfnamefont {P.~J.}\ \bibnamefont {Hellyer}}, \bibinfo {author}
  {\bibfnamefont {N.}~\bibnamefont {Mirzaei}}, \bibinfo {author} {\bibfnamefont
  {C.}~\bibnamefont {Leeson}}, \bibinfo {author} {\bibfnamefont {D.~W.}\
  \bibnamefont {Carmichael}}, \bibinfo {author} {\bibfnamefont {D.~J.}\
  \bibnamefont {Sharp}}, \bibinfo {author} {\bibfnamefont {W.~L.}\ \bibnamefont
  {Shew}},\ and\ \bibinfo {author} {\bibfnamefont {R.}~\bibnamefont {Leech}},\
  }\bibfield  {title} {\bibinfo {title} {Cascades and {{Cognitive State}}:
  {{Focused Attention Incurs Subcritical Dynamics}}},\ }\href
  {https://doi.org/10.1523/JNEUROSCI.3694-14.2015} {\bibfield  {journal}
  {\bibinfo  {journal} {Journal of Neuroscience}\ }\textbf {\bibinfo {volume}
  {35}},\ \bibinfo {pages} {4626} (\bibinfo {year} {2015})}\BibitemShut
  {NoStop}%
\bibitem [{\citenamefont {Kinouchi}\ and\ \citenamefont
  {Copelli}(2006)}]{kinouchiOptimalDynamicalRange2006}%
  \BibitemOpen
  \bibfield  {author} {\bibinfo {author} {\bibfnamefont {O.}~\bibnamefont
  {Kinouchi}}\ and\ \bibinfo {author} {\bibfnamefont {M.}~\bibnamefont
  {Copelli}},\ }\bibfield  {title} {\bibinfo {title} {Optimal dynamical range
  of excitable networks at criticality},\ }\href
  {https://doi.org/10.1038/nphys289} {\bibfield  {journal} {\bibinfo  {journal}
  {Nature Physics}\ }\textbf {\bibinfo {volume} {2}},\ \bibinfo {pages} {348}
  (\bibinfo {year} {2006})}\BibitemShut {NoStop}%
\bibitem [{\citenamefont {Shew}\ \emph {et~al.}(2009)\citenamefont {Shew},
  \citenamefont {Yang}, \citenamefont {Petermann}, \citenamefont {Roy},\ and\
  \citenamefont {Plenz}}]{shewNeuronalAvalanchesImply2009}%
  \BibitemOpen
  \bibfield  {author} {\bibinfo {author} {\bibfnamefont {W.~L.}\ \bibnamefont
  {Shew}}, \bibinfo {author} {\bibfnamefont {H.}~\bibnamefont {Yang}}, \bibinfo
  {author} {\bibfnamefont {T.}~\bibnamefont {Petermann}}, \bibinfo {author}
  {\bibfnamefont {R.}~\bibnamefont {Roy}},\ and\ \bibinfo {author}
  {\bibfnamefont {D.}~\bibnamefont {Plenz}},\ }\bibfield  {title} {\bibinfo
  {title} {Neuronal avalanches imply maximum dynamic range in cortical networks
  at criticality},\ }\href {https://doi.org/10.1523/JNEUROSCI.3864-09.2009}
  {\bibfield  {journal} {\bibinfo  {journal} {The Journal of Neuroscience: The
  Official Journal of the Society for Neuroscience}\ }\textbf {\bibinfo
  {volume} {29}},\ \bibinfo {pages} {15595} (\bibinfo {year}
  {2009})}\BibitemShut {NoStop}%
\bibitem [{\citenamefont {Gautam}\ \emph {et~al.}(2015)\citenamefont {Gautam},
  \citenamefont {Hoang}, \citenamefont {McClanahan}, \citenamefont {Grady},\
  and\ \citenamefont {Shew}}]{gautamMaximizingSensoryDynamic2015}%
  \BibitemOpen
  \bibfield  {author} {\bibinfo {author} {\bibfnamefont {S.~H.}\ \bibnamefont
  {Gautam}}, \bibinfo {author} {\bibfnamefont {T.~T.}\ \bibnamefont {Hoang}},
  \bibinfo {author} {\bibfnamefont {K.}~\bibnamefont {McClanahan}}, \bibinfo
  {author} {\bibfnamefont {S.~K.}\ \bibnamefont {Grady}},\ and\ \bibinfo
  {author} {\bibfnamefont {W.~L.}\ \bibnamefont {Shew}},\ }\bibfield  {title}
  {\bibinfo {title} {Maximizing {{Sensory Dynamic Range}} by {{Tuning}} the
  {{Cortical State}} to {{Criticality}}},\ }\href
  {https://doi.org/10.1371/journal.pcbi.1004576} {\bibfield  {journal}
  {\bibinfo  {journal} {PLOS Computational Biology}\ }\textbf {\bibinfo
  {volume} {11}},\ \bibinfo {pages} {e1004576} (\bibinfo {year}
  {2015})}\BibitemShut {NoStop}%
\bibitem [{\citenamefont {Beggs}(2022)}]{beggsCortexCriticalPoint2022}%
  \BibitemOpen
  \bibfield  {author} {\bibinfo {author} {\bibfnamefont {J.~M.}\ \bibnamefont
  {Beggs}},\ }\href {https://doi.org/10.7551/mitpress/13588.001.0001} {\emph
  {\bibinfo {title} {The {{Cortex}} and the {{Critical Point}}:
  {{Understanding}} the {{Power}} of {{Emergence}}}}}\ (\bibinfo  {publisher}
  {The MIT Press},\ \bibinfo {year} {2022})\BibitemShut {NoStop}%
\bibitem [{\citenamefont {Vock}\ and\ \citenamefont
  {Meisel}()}]{vockCriticalDynamicsGoverns2026}%
  \BibitemOpen
  \bibfield  {author} {\bibinfo {author} {\bibfnamefont {S.}~\bibnamefont
  {Vock}}\ and\ \bibinfo {author} {\bibfnamefont {C.}~\bibnamefont {Meisel}},\
  }\href {https://doi.org/10.48550/arXiv.2507.08527} {\bibinfo {title}
  {Critical dynamics governs deep learning}},\ \Eprint
  {https://arxiv.org/abs/2507.08527} {2507.08527} \BibitemShut {NoStop}%
\bibitem [{\citenamefont {Sun}\ \emph {et~al.}(2025)\citenamefont {Sun},
  \citenamefont {Sipling}, \citenamefont {Zhang},\ and\ \citenamefont
  {Di~Ventra}}]{sunMemoryNeuralActivity2025}%
  \BibitemOpen
  \bibfield  {author} {\bibinfo {author} {\bibfnamefont {J.~K.-C.}\
  \bibnamefont {Sun}}, \bibinfo {author} {\bibfnamefont {C.}~\bibnamefont
  {Sipling}}, \bibinfo {author} {\bibfnamefont {Y.-H.}\ \bibnamefont {Zhang}},\
  and\ \bibinfo {author} {\bibfnamefont {M.}~\bibnamefont {Di~Ventra}},\
  }\bibfield  {title} {\bibinfo {title} {Memory in neural activity:
  {{Long-range}} order without criticality},\ }\href
  {https://doi.org/10.1103/d7mp-pk1w} {\bibfield  {journal} {\bibinfo
  {journal} {Physical Review E}\ }\textbf {\bibinfo {volume} {112}},\ \bibinfo
  {pages} {064401} (\bibinfo {year} {2025})}\BibitemShut {NoStop}%
\bibitem [{\citenamefont {Sipling}\ \emph {et~al.}(2026)\citenamefont
  {Sipling}, \citenamefont {Zhang},\ and\ \citenamefont {{Di
  Ventra}}}]{siplingCriticalAssessmentBrain2026}%
  \BibitemOpen
  \bibfield  {author} {\bibinfo {author} {\bibfnamefont {C.}~\bibnamefont
  {Sipling}}, \bibinfo {author} {\bibfnamefont {Y.-H.}\ \bibnamefont {Zhang}},\
  and\ \bibinfo {author} {\bibfnamefont {M.}~\bibnamefont {{Di Ventra}}},\
  }\bibfield  {title} {\bibinfo {title} {A critical assessment of the brain
  criticality hypothesis},\ }\bibfield  {journal} {\bibinfo  {journal} {Trends
  Open}\ }\href {https://doi.org/https://doi.org/10.1016/j.treopn.2026.06.001}
  {https://doi.org/10.1016/j.treopn.2026.06.001} (\bibinfo {year}
  {2026})\BibitemShut {NoStop}%
\bibitem [{\citenamefont {Roberts}\ \emph {et~al.}(2014)\citenamefont
  {Roberts}, \citenamefont {Iyer}, \citenamefont {Vanhatalo},\ and\
  \citenamefont {Breakspear}}]{robertsCriticalRoleResource2014}%
  \BibitemOpen
  \bibfield  {author} {\bibinfo {author} {\bibfnamefont {J.~A.}\ \bibnamefont
  {Roberts}}, \bibinfo {author} {\bibfnamefont {K.~K.}\ \bibnamefont {Iyer}},
  \bibinfo {author} {\bibfnamefont {S.}~\bibnamefont {Vanhatalo}},\ and\
  \bibinfo {author} {\bibfnamefont {M.}~\bibnamefont {Breakspear}},\ }\bibfield
   {title} {\bibinfo {title} {Critical role for resource constraints in neural
  models},\ }\bibfield  {journal} {\bibinfo  {journal} {Frontiers in Systems
  Neuroscience}\ }\textbf {\bibinfo {volume} {8}},\ \href
  {https://doi.org/10.3389/fnsys.2014.00154} {10.3389/fnsys.2014.00154}
  (\bibinfo {year} {2014})\BibitemShut {NoStop}%
\bibitem [{\citenamefont {Virkar}\ \emph {et~al.}(2016)\citenamefont {Virkar},
  \citenamefont {Shew}, \citenamefont {Restrepo},\ and\ \citenamefont
  {Ott}}]{virkarFeedbackControlStabilization2016}%
  \BibitemOpen
  \bibfield  {author} {\bibinfo {author} {\bibfnamefont {Y.~S.}\ \bibnamefont
  {Virkar}}, \bibinfo {author} {\bibfnamefont {W.~L.}\ \bibnamefont {Shew}},
  \bibinfo {author} {\bibfnamefont {J.~G.}\ \bibnamefont {Restrepo}},\ and\
  \bibinfo {author} {\bibfnamefont {E.}~\bibnamefont {Ott}},\ }\bibfield
  {title} {\bibinfo {title} {Feedback control stabilization of critical
  dynamics via resource transport on multilayer networks: {{How}} glia enable
  learning dynamics in the brain},\ }\href
  {https://doi.org/10.1103/PhysRevE.94.042310} {\bibfield  {journal} {\bibinfo
  {journal} {Physical Review E}\ }\textbf {\bibinfo {volume} {94}},\ \bibinfo
  {pages} {042310} (\bibinfo {year} {2016})}\BibitemShut {NoStop}%
\bibitem [{\citenamefont {Franovi{\'c}}\ \emph {et~al.}(2022)\citenamefont
  {Franovi{\'c}}, \citenamefont {Eydam}, \citenamefont {Yanchuk},\ and\
  \citenamefont {Berner}}]{franovicCollectiveActivityBursting2022}%
  \BibitemOpen
  \bibfield  {author} {\bibinfo {author} {\bibfnamefont {I.}~\bibnamefont
  {Franovi{\'c}}}, \bibinfo {author} {\bibfnamefont {S.}~\bibnamefont {Eydam}},
  \bibinfo {author} {\bibfnamefont {S.}~\bibnamefont {Yanchuk}},\ and\ \bibinfo
  {author} {\bibfnamefont {R.}~\bibnamefont {Berner}},\ }\bibfield  {title}
  {\bibinfo {title} {Collective {{Activity Bursting}} in a {{Population}} of
  {{Excitable Units Adaptively Coupled}} to a {{Pool}} of {{Resources}}},\
  }\bibfield  {journal} {\bibinfo  {journal} {Frontiers in Network Physiology}\
  }\textbf {\bibinfo {volume} {2}},\ \href
  {https://doi.org/10.3389/fnetp.2022.841829} {10.3389/fnetp.2022.841829}
  (\bibinfo {year} {2022})\BibitemShut {NoStop}%
\bibitem [{\citenamefont {Kinouchi}\ \emph {et~al.}(2019)\citenamefont
  {Kinouchi}, \citenamefont {Brochini}, \citenamefont {Costa}, \citenamefont
  {Campos},\ and\ \citenamefont
  {Copelli}}]{kinouchiStochasticOscillationsDragon2019}%
  \BibitemOpen
  \bibfield  {author} {\bibinfo {author} {\bibfnamefont {O.}~\bibnamefont
  {Kinouchi}}, \bibinfo {author} {\bibfnamefont {L.}~\bibnamefont {Brochini}},
  \bibinfo {author} {\bibfnamefont {A.~A.}\ \bibnamefont {Costa}}, \bibinfo
  {author} {\bibfnamefont {J.~G.~F.}\ \bibnamefont {Campos}},\ and\ \bibinfo
  {author} {\bibfnamefont {M.}~\bibnamefont {Copelli}},\ }\bibfield  {title}
  {\bibinfo {title} {Stochastic oscillations and dragon king avalanches in
  self-organized quasi-critical systems},\ }\href
  {https://doi.org/10.1038/s41598-019-40473-1} {\bibfield  {journal} {\bibinfo
  {journal} {Scientific Reports}\ }\textbf {\bibinfo {volume} {9}},\ \bibinfo
  {pages} {3874} (\bibinfo {year} {2019})}\BibitemShut {NoStop}%
\bibitem [{\citenamefont {{de
  Arcangelis}}(2012)}]{dearcangelisAreDragonkingNeuronal2012}%
  \BibitemOpen
  \bibfield  {author} {\bibinfo {author} {\bibfnamefont {L.}~\bibnamefont {{de
  Arcangelis}}},\ }\bibfield  {title} {\bibinfo {title} {Are dragon-king
  neuronal avalanches dungeons for self-organized brain activity?},\ }\href
  {https://doi.org/10.1140/epjst/e2012-01574-6} {\bibfield  {journal} {\bibinfo
   {journal} {The European Physical Journal Special Topics}\ }\textbf {\bibinfo
  {volume} {205}},\ \bibinfo {pages} {243} (\bibinfo {year}
  {2012})}\BibitemShut {NoStop}%
\bibitem [{\citenamefont {Mishra}\ \emph {et~al.}(2018)\citenamefont {Mishra},
  \citenamefont {Saha}, \citenamefont {Vigneshwaran}, \citenamefont {Pal},
  \citenamefont {Kapitaniak},\ and\ \citenamefont
  {Dana}}]{mishraDragonkinglikeExtremeEvents2018}%
  \BibitemOpen
  \bibfield  {author} {\bibinfo {author} {\bibfnamefont {A.}~\bibnamefont
  {Mishra}}, \bibinfo {author} {\bibfnamefont {S.}~\bibnamefont {Saha}},
  \bibinfo {author} {\bibfnamefont {M.}~\bibnamefont {Vigneshwaran}}, \bibinfo
  {author} {\bibfnamefont {P.}~\bibnamefont {Pal}}, \bibinfo {author}
  {\bibfnamefont {T.}~\bibnamefont {Kapitaniak}},\ and\ \bibinfo {author}
  {\bibfnamefont {S.~K.}\ \bibnamefont {Dana}},\ }\bibfield  {title} {\bibinfo
  {title} {Dragon-king-like extreme events in coupled bursting neurons},\
  }\href {https://doi.org/10.1103/PhysRevE.97.062311} {\bibfield  {journal}
  {\bibinfo  {journal} {Physical Review E}\ }\textbf {\bibinfo {volume} {97}},\
  \bibinfo {pages} {062311} (\bibinfo {year} {2018})}\BibitemShut {NoStop}%
\bibitem [{\citenamefont {Ates}\ \emph {et~al.}(2007)\citenamefont {Ates},
  \citenamefont {Pohl}, \citenamefont {Pattard},\ and\ \citenamefont
  {Rost}}]{atesAntiblockadeRydbergExcitation2007}%
  \BibitemOpen
  \bibfield  {author} {\bibinfo {author} {\bibfnamefont {C.}~\bibnamefont
  {Ates}}, \bibinfo {author} {\bibfnamefont {T.}~\bibnamefont {Pohl}}, \bibinfo
  {author} {\bibfnamefont {T.}~\bibnamefont {Pattard}},\ and\ \bibinfo {author}
  {\bibfnamefont {J.~M.}\ \bibnamefont {Rost}},\ }\bibfield  {title} {\bibinfo
  {title} {Antiblockade in {{Rydberg Excitation}} of an {{Ultracold Lattice
  Gas}}},\ }\href {https://doi.org/10.1103/PhysRevLett.98.023002} {\bibfield
  {journal} {\bibinfo  {journal} {Physical Review Letters}\ }\textbf {\bibinfo
  {volume} {98}},\ \bibinfo {pages} {023002} (\bibinfo {year}
  {2007})}\BibitemShut {NoStop}%
\bibitem [{\citenamefont {Amthor}\ \emph {et~al.}(2010)\citenamefont {Amthor},
  \citenamefont {Giese}, \citenamefont {Hofmann},\ and\ \citenamefont
  {Weidem{\"u}ller}}]{amthorEvidenceAntiblockadeUltracold2010}%
  \BibitemOpen
  \bibfield  {author} {\bibinfo {author} {\bibfnamefont {T.}~\bibnamefont
  {Amthor}}, \bibinfo {author} {\bibfnamefont {C.}~\bibnamefont {Giese}},
  \bibinfo {author} {\bibfnamefont {C.~S.}\ \bibnamefont {Hofmann}},\ and\
  \bibinfo {author} {\bibfnamefont {M.}~\bibnamefont {Weidem{\"u}ller}},\
  }\bibfield  {title} {\bibinfo {title} {Evidence of {{Antiblockade}} in an
  {{Ultracold Rydberg Gas}}},\ }\href
  {https://doi.org/10.1103/PhysRevLett.104.013001} {\bibfield  {journal}
  {\bibinfo  {journal} {Physical Review Letters}\ }\textbf {\bibinfo {volume}
  {104}},\ \bibinfo {pages} {013001} (\bibinfo {year} {2010})}\BibitemShut
  {NoStop}%
\bibitem [{\citenamefont {Rutten}\ and\ \citenamefont
  {Sanders}(2021)}]{ruttenModelingRydbergGases2021}%
  \BibitemOpen
  \bibfield  {author} {\bibinfo {author} {\bibfnamefont {D.}~\bibnamefont
  {Rutten}}\ and\ \bibinfo {author} {\bibfnamefont {J.}~\bibnamefont
  {Sanders}},\ }\bibfield  {title} {\bibinfo {title} {Modeling {{Rydberg}}
  gases using random sequential adsorption on random graphs},\ }\href
  {https://doi.org/10.1103/PhysRevA.103.033302} {\bibfield  {journal} {\bibinfo
   {journal} {Physical Review A}\ }\textbf {\bibinfo {volume} {103}},\ \bibinfo
  {pages} {033302} (\bibinfo {year} {2021})}\BibitemShut {NoStop}%
\bibitem [{\citenamefont {Ohler}\ \emph {et~al.}(2025)\citenamefont {Ohler},
  \citenamefont {Brady}, \citenamefont {Mischke}, \citenamefont {Bender},
  \citenamefont {Ott}, \citenamefont {Niederpr{\"u}m}, \citenamefont {Ripken},
  \citenamefont {Otterbach},\ and\ \citenamefont
  {Fleischhauer}}]{ohlerNonequilibriumUniversalityRydbergexcitation2025}%
  \BibitemOpen
  \bibfield  {author} {\bibinfo {author} {\bibfnamefont {S.}~\bibnamefont
  {Ohler}}, \bibinfo {author} {\bibfnamefont {D.}~\bibnamefont {Brady}},
  \bibinfo {author} {\bibfnamefont {P.}~\bibnamefont {Mischke}}, \bibinfo
  {author} {\bibfnamefont {J.}~\bibnamefont {Bender}}, \bibinfo {author}
  {\bibfnamefont {H.}~\bibnamefont {Ott}}, \bibinfo {author} {\bibfnamefont
  {T.}~\bibnamefont {Niederpr{\"u}m}}, \bibinfo {author} {\bibfnamefont
  {W.}~\bibnamefont {Ripken}}, \bibinfo {author} {\bibfnamefont {J.~S.}\
  \bibnamefont {Otterbach}},\ and\ \bibinfo {author} {\bibfnamefont
  {M.}~\bibnamefont {Fleischhauer}},\ }\bibfield  {title} {\bibinfo {title}
  {Nonequilibrium universality of {{Rydberg-excitation}} spreading on a dynamic
  network},\ }\href {https://doi.org/10.1103/8rlg-169g} {\bibfield  {journal}
  {\bibinfo  {journal} {Physical Review Research}\ }\textbf {\bibinfo {volume}
  {7}},\ \bibinfo {pages} {033167} (\bibinfo {year} {2025})}\BibitemShut
  {NoStop}%
\bibitem [{\citenamefont {Levi}\ \emph {et~al.}(2016)\citenamefont {Levi},
  \citenamefont {Guti{\'e}rrez},\ and\ \citenamefont
  {Lesanovsky}}]{leviQuantumNonequilibriumDynamics2016}%
  \BibitemOpen
  \bibfield  {author} {\bibinfo {author} {\bibfnamefont {E.}~\bibnamefont
  {Levi}}, \bibinfo {author} {\bibfnamefont {R.}~\bibnamefont
  {Guti{\'e}rrez}},\ and\ \bibinfo {author} {\bibfnamefont {I.}~\bibnamefont
  {Lesanovsky}},\ }\bibfield  {title} {\bibinfo {title} {Quantum
  non-equilibrium dynamics of {{Rydberg}} gases in the presence of dephasing
  noise of different strengths},\ }\href
  {https://doi.org/10.1088/0953-4075/49/18/184003} {\bibfield  {journal}
  {\bibinfo  {journal} {Journal of Physics B: Atomic, Molecular and Optical
  Physics}\ }\textbf {\bibinfo {volume} {49}},\ \bibinfo {pages} {184003}
  (\bibinfo {year} {2016})}\BibitemShut {NoStop}%
\bibitem [{\citenamefont {Schempp}\ \emph {et~al.}(2014)\citenamefont
  {Schempp}, \citenamefont {G{\"u}nter}, \citenamefont
  {{Robert-de-Saint-Vincent}}, \citenamefont {Hofmann}, \citenamefont {Breyel},
  \citenamefont {Komnik}, \citenamefont {Sch{\"o}nleber}, \citenamefont
  {G{\"a}rttner}, \citenamefont {Evers}, \citenamefont {Whitlock},\ and\
  \citenamefont {Weidem{\"u}ller}}]{schemppFullCountingStatistics2014a}%
  \BibitemOpen
  \bibfield  {author} {\bibinfo {author} {\bibfnamefont {H.}~\bibnamefont
  {Schempp}}, \bibinfo {author} {\bibfnamefont {G.}~\bibnamefont {G{\"u}nter}},
  \bibinfo {author} {\bibfnamefont {M.}~\bibnamefont
  {{Robert-de-Saint-Vincent}}}, \bibinfo {author} {\bibfnamefont {C.~S.}\
  \bibnamefont {Hofmann}}, \bibinfo {author} {\bibfnamefont {D.}~\bibnamefont
  {Breyel}}, \bibinfo {author} {\bibfnamefont {A.}~\bibnamefont {Komnik}},
  \bibinfo {author} {\bibfnamefont {D.~W.}\ \bibnamefont {Sch{\"o}nleber}},
  \bibinfo {author} {\bibfnamefont {M.}~\bibnamefont {G{\"a}rttner}}, \bibinfo
  {author} {\bibfnamefont {J.}~\bibnamefont {Evers}}, \bibinfo {author}
  {\bibfnamefont {S.}~\bibnamefont {Whitlock}},\ and\ \bibinfo {author}
  {\bibfnamefont {M.}~\bibnamefont {Weidem{\"u}ller}},\ }\bibfield  {title}
  {\bibinfo {title} {Full {{Counting Statistics}} of {{Laser Excited Rydberg
  Aggregates}} in a {{One-Dimensional Geometry}}},\ }\href
  {https://doi.org/10.1103/PhysRevLett.112.013002} {\bibfield  {journal}
  {\bibinfo  {journal} {Physical Review Letters}\ }\textbf {\bibinfo {volume}
  {112}},\ \bibinfo {pages} {013002} (\bibinfo {year} {2014})}\BibitemShut
  {NoStop}%
\bibitem [{\citenamefont {Malossi}\ \emph {et~al.}(2014)\citenamefont
  {Malossi}, \citenamefont {Valado}, \citenamefont {Scotto}, \citenamefont
  {Huillery}, \citenamefont {Pillet}, \citenamefont {Ciampini}, \citenamefont
  {Arimondo},\ and\ \citenamefont
  {Morsch}}]{malossiFullCountingStatistics2014}%
  \BibitemOpen
  \bibfield  {author} {\bibinfo {author} {\bibfnamefont {N.}~\bibnamefont
  {Malossi}}, \bibinfo {author} {\bibfnamefont {M.~M.}\ \bibnamefont {Valado}},
  \bibinfo {author} {\bibfnamefont {S.}~\bibnamefont {Scotto}}, \bibinfo
  {author} {\bibfnamefont {P.}~\bibnamefont {Huillery}}, \bibinfo {author}
  {\bibfnamefont {P.}~\bibnamefont {Pillet}}, \bibinfo {author} {\bibfnamefont
  {D.}~\bibnamefont {Ciampini}}, \bibinfo {author} {\bibfnamefont
  {E.}~\bibnamefont {Arimondo}},\ and\ \bibinfo {author} {\bibfnamefont
  {O.}~\bibnamefont {Morsch}},\ }\bibfield  {title} {\bibinfo {title} {Full
  {{Counting Statistics}} and {{Phase Diagram}} of a {{Dissipative Rydberg
  Gas}}},\ }\href {https://doi.org/10.1103/PhysRevLett.113.023006} {\bibfield
  {journal} {\bibinfo  {journal} {Physical Review Letters}\ }\textbf {\bibinfo
  {volume} {113}},\ \bibinfo {pages} {023006} (\bibinfo {year}
  {2014})}\BibitemShut {NoStop}%
\bibitem [{\citenamefont {Marcuzzi}\ \emph {et~al.}(2015)\citenamefont
  {Marcuzzi}, \citenamefont {Levi}, \citenamefont {Li}, \citenamefont
  {Garrahan}, \citenamefont {Olmos},\ and\ \citenamefont
  {Lesanovsky}}]{marcuzziNonequilibriumUniversalityDynamics2015}%
  \BibitemOpen
  \bibfield  {author} {\bibinfo {author} {\bibfnamefont {M.}~\bibnamefont
  {Marcuzzi}}, \bibinfo {author} {\bibfnamefont {E.}~\bibnamefont {Levi}},
  \bibinfo {author} {\bibfnamefont {W.}~\bibnamefont {Li}}, \bibinfo {author}
  {\bibfnamefont {J.~P.}\ \bibnamefont {Garrahan}}, \bibinfo {author}
  {\bibfnamefont {B.}~\bibnamefont {Olmos}},\ and\ \bibinfo {author}
  {\bibfnamefont {I.}~\bibnamefont {Lesanovsky}},\ }\bibfield  {title}
  {\bibinfo {title} {Non-equilibrium universality in the dynamics of
  dissipative cold atomic gases},\ }\href
  {https://doi.org/10.1088/1367-2630/17/7/072003} {\bibfield  {journal}
  {\bibinfo  {journal} {New Journal of Physics}\ }\textbf {\bibinfo {volume}
  {17}},\ \bibinfo {pages} {072003} (\bibinfo {year} {2015})}\BibitemShut
  {NoStop}%
\bibitem [{\citenamefont {Helmrich}\ \emph {et~al.}(2020)\citenamefont
  {Helmrich}, \citenamefont {Arias}, \citenamefont {Lochead}, \citenamefont
  {Wintermantel}, \citenamefont {Buchhold}, \citenamefont {Diehl},\ and\
  \citenamefont {Whitlock}}]{helmrichSignaturesSelforganizedCriticality2020}%
  \BibitemOpen
  \bibfield  {author} {\bibinfo {author} {\bibfnamefont {S.}~\bibnamefont
  {Helmrich}}, \bibinfo {author} {\bibfnamefont {A.}~\bibnamefont {Arias}},
  \bibinfo {author} {\bibfnamefont {G.}~\bibnamefont {Lochead}}, \bibinfo
  {author} {\bibfnamefont {T.~M.}\ \bibnamefont {Wintermantel}}, \bibinfo
  {author} {\bibfnamefont {M.}~\bibnamefont {Buchhold}}, \bibinfo {author}
  {\bibfnamefont {S.}~\bibnamefont {Diehl}},\ and\ \bibinfo {author}
  {\bibfnamefont {S.}~\bibnamefont {Whitlock}},\ }\bibfield  {title} {\bibinfo
  {title} {Signatures of self-organized criticality in an ultracold atomic
  gas},\ }\href {https://doi.org/10.1038/s41586-019-1908-6} {\bibfield
  {journal} {\bibinfo  {journal} {Nature}\ }\textbf {\bibinfo {volume} {577}},\
  \bibinfo {pages} {481} (\bibinfo {year} {2020})}\BibitemShut {NoStop}%
\bibitem [{\citenamefont {Wintermantel}\ \emph {et~al.}(2021)\citenamefont
  {Wintermantel}, \citenamefont {Buchhold}, \citenamefont {Shevate},
  \citenamefont {Morgado}, \citenamefont {Wang}, \citenamefont {Lochead},
  \citenamefont {Diehl},\ and\ \citenamefont
  {Whitlock}}]{wintermantelEpidemicGrowthGriffiths2021}%
  \BibitemOpen
  \bibfield  {author} {\bibinfo {author} {\bibfnamefont {T.~M.}\ \bibnamefont
  {Wintermantel}}, \bibinfo {author} {\bibfnamefont {M.}~\bibnamefont
  {Buchhold}}, \bibinfo {author} {\bibfnamefont {S.}~\bibnamefont {Shevate}},
  \bibinfo {author} {\bibfnamefont {M.}~\bibnamefont {Morgado}}, \bibinfo
  {author} {\bibfnamefont {Y.}~\bibnamefont {Wang}}, \bibinfo {author}
  {\bibfnamefont {G.}~\bibnamefont {Lochead}}, \bibinfo {author} {\bibfnamefont
  {S.}~\bibnamefont {Diehl}},\ and\ \bibinfo {author} {\bibfnamefont
  {S.}~\bibnamefont {Whitlock}},\ }\bibfield  {title} {\bibinfo {title}
  {Epidemic growth and {{Griffiths}} effects on an emergent network of excited
  atoms},\ }\href {https://doi.org/10.1038/s41467-020-20333-7} {\bibfield
  {journal} {\bibinfo  {journal} {Nature Communications}\ }\textbf {\bibinfo
  {volume} {12}},\ \bibinfo {pages} {103} (\bibinfo {year} {2021})}\BibitemShut
  {NoStop}%
\bibitem [{\citenamefont {Klocke}\ \emph {et~al.}(2021)\citenamefont {Klocke},
  \citenamefont {Wintermantel}, \citenamefont {Lochead}, \citenamefont
  {Whitlock},\ and\ \citenamefont
  {Buchhold}}]{klockeHydrodynamicStabilizationSelfOrganized2021}%
  \BibitemOpen
  \bibfield  {author} {\bibinfo {author} {\bibfnamefont {K.}~\bibnamefont
  {Klocke}}, \bibinfo {author} {\bibfnamefont {T.~M.}\ \bibnamefont
  {Wintermantel}}, \bibinfo {author} {\bibfnamefont {G.}~\bibnamefont
  {Lochead}}, \bibinfo {author} {\bibfnamefont {S.}~\bibnamefont {Whitlock}},\
  and\ \bibinfo {author} {\bibfnamefont {M.}~\bibnamefont {Buchhold}},\
  }\bibfield  {title} {\bibinfo {title} {Hydrodynamic {{Stabilization}} of
  {{Self-Organized Criticality}} in a {{Driven Rydberg Gas}}},\ }\href
  {https://doi.org/10.1103/PhysRevLett.126.123401} {\bibfield  {journal}
  {\bibinfo  {journal} {Physical Review Letters}\ }\textbf {\bibinfo {volume}
  {126}},\ \bibinfo {pages} {123401} (\bibinfo {year} {2021})}\BibitemShut
  {NoStop}%
\bibitem [{\citenamefont {Virkar}(2020)}]{virkarDynamicRegulationResource2020}%
  \BibitemOpen
  \bibfield  {author} {\bibinfo {author} {\bibfnamefont {Y.~S.}\ \bibnamefont
  {Virkar}},\ }\bibfield  {title} {\bibinfo {title} {Dynamic regulation of
  resource transport induces criticality in interdependent networks of
  excitable units},\ }\bibfield  {journal} {\bibinfo  {journal} {Physical
  Review E}\ }\textbf {\bibinfo {volume} {101}},\ \href
  {https://doi.org/10.1103/PhysRevE.101.022303} {10.1103/PhysRevE.101.022303}
  (\bibinfo {year} {2020})\BibitemShut {NoStop}%
\bibitem [{\citenamefont {Sethna}\ \emph {et~al.}(2001)\citenamefont {Sethna},
  \citenamefont {Dahmen},\ and\ \citenamefont
  {Myers}}]{sethnaCracklingNoise2001}%
  \BibitemOpen
  \bibfield  {author} {\bibinfo {author} {\bibfnamefont {J.~P.}\ \bibnamefont
  {Sethna}}, \bibinfo {author} {\bibfnamefont {K.~A.}\ \bibnamefont {Dahmen}},\
  and\ \bibinfo {author} {\bibfnamefont {C.~R.}\ \bibnamefont {Myers}},\
  }\bibfield  {title} {\bibinfo {title} {Crackling noise},\ }\href
  {https://doi.org/10.1038/35065675} {\bibfield  {journal} {\bibinfo  {journal}
  {Nature}\ }\textbf {\bibinfo {volume} {410}},\ \bibinfo {pages} {242}
  (\bibinfo {year} {2001})}\BibitemShut {NoStop}%
\bibitem [{\citenamefont {Markovi{\'c}}\ and\ \citenamefont
  {Gros}(2014)}]{markovicPowerLawsSelforganized2014}%
  \BibitemOpen
  \bibfield  {author} {\bibinfo {author} {\bibfnamefont {D.}~\bibnamefont
  {Markovi{\'c}}}\ and\ \bibinfo {author} {\bibfnamefont {C.}~\bibnamefont
  {Gros}},\ }\bibfield  {title} {\bibinfo {title} {Power laws and
  self-organized criticality in theory and nature},\ }\href
  {https://doi.org/10.1016/j.physrep.2013.11.002} {\bibfield  {journal}
  {\bibinfo  {journal} {Physics Reports}\ }\bibinfo {series} {Power Laws and
  {{Self-Organized Criticality}} in {{Theory}} and {{Nature}}},\ \textbf
  {\bibinfo {volume} {536}},\ \bibinfo {pages} {41} (\bibinfo {year}
  {2014})}\BibitemShut {NoStop}%
\bibitem [{\citenamefont {Brady}\ \emph {et~al.}(2024)\citenamefont {Brady},
  \citenamefont {Ohler}, \citenamefont {Otterbach},\ and\ \citenamefont
  {Fleischhauer}}]{bradyAnomalousDirectedPercolation2024}%
  \BibitemOpen
  \bibfield  {author} {\bibinfo {author} {\bibfnamefont {D.}~\bibnamefont
  {Brady}}, \bibinfo {author} {\bibfnamefont {S.}~\bibnamefont {Ohler}},
  \bibinfo {author} {\bibfnamefont {J.}~\bibnamefont {Otterbach}},\ and\
  \bibinfo {author} {\bibfnamefont {M.}~\bibnamefont {Fleischhauer}},\
  }\bibfield  {title} {\bibinfo {title} {Anomalous {{Directed Percolation}} on
  a {{Dynamic Network Using Rydberg Facilitation}}},\ }\href
  {https://doi.org/10.1103/PhysRevLett.133.173401} {\bibfield  {journal}
  {\bibinfo  {journal} {Physical Review Letters}\ }\textbf {\bibinfo {volume}
  {133}},\ \bibinfo {pages} {173401} (\bibinfo {year} {2024})}\BibitemShut
  {NoStop}%
\bibitem [{\citenamefont {Fosque}\ \emph {et~al.}(2021)\citenamefont {Fosque},
  \citenamefont {Williams-Garc\'{\i}a}, \citenamefont {Beggs},\ and\
  \citenamefont {Ortiz}}]{fosqueEvidenceQuasicriticalBrain2021}%
  \BibitemOpen
  \bibfield  {author} {\bibinfo {author} {\bibfnamefont {L.~J.}\ \bibnamefont
  {Fosque}}, \bibinfo {author} {\bibfnamefont {R.~V.}\ \bibnamefont
  {Williams-Garc\'{\i}a}}, \bibinfo {author} {\bibfnamefont {J.~M.}\
  \bibnamefont {Beggs}},\ and\ \bibinfo {author} {\bibfnamefont
  {G.}~\bibnamefont {Ortiz}},\ }\bibfield  {title} {\bibinfo {title} {Evidence
  for quasicritical brain dynamics},\ }\href
  {https://doi.org/10.1103/PhysRevLett.126.098101} {\bibfield  {journal}
  {\bibinfo  {journal} {Phys. Rev. Lett.}\ }\textbf {\bibinfo {volume} {126}},\
  \bibinfo {pages} {098101} (\bibinfo {year} {2021})}\BibitemShut {NoStop}%
\bibitem [{\citenamefont {Spasojevi{\'c}}\ \emph {et~al.}(1996)\citenamefont
  {Spasojevi{\'c}}, \citenamefont {Bukvi{\'c}}, \citenamefont {Milo{\v
  s}evi{\'c}},\ and\ \citenamefont
  {Stanley}}]{spasojevicBarkhausenNoiseElementary1996}%
  \BibitemOpen
  \bibfield  {author} {\bibinfo {author} {\bibfnamefont {D.}~\bibnamefont
  {Spasojevi{\'c}}}, \bibinfo {author} {\bibfnamefont {S.}~\bibnamefont
  {Bukvi{\'c}}}, \bibinfo {author} {\bibfnamefont {S.}~\bibnamefont {Milo{\v
  s}evi{\'c}}},\ and\ \bibinfo {author} {\bibfnamefont {H.~E.}\ \bibnamefont
  {Stanley}},\ }\bibfield  {title} {\bibinfo {title} {Barkhausen noise:
  {{Elementary}} signals, power laws, and scaling relations},\ }\href
  {https://doi.org/10.1103/PhysRevE.54.2531} {\bibfield  {journal} {\bibinfo
  {journal} {Physical Review E}\ }\textbf {\bibinfo {volume} {54}},\ \bibinfo
  {pages} {2531} (\bibinfo {year} {1996})}\BibitemShut {NoStop}%
\bibitem [{\citenamefont
  {Fries}(2015)}]{friesRhythmsCognitionCommunication2015}%
  \BibitemOpen
  \bibfield  {author} {\bibinfo {author} {\bibfnamefont {P.}~\bibnamefont
  {Fries}},\ }\bibfield  {title} {\bibinfo {title} {Rhythms for {{Cognition}}:
  {{Communication}} through {{Coherence}}},\ }\href
  {https://doi.org/10.1016/j.neuron.2015.09.034} {\bibfield  {journal}
  {\bibinfo  {journal} {Neuron}\ }\textbf {\bibinfo {volume} {88}},\ \bibinfo
  {pages} {220} (\bibinfo {year} {2015})}\BibitemShut {NoStop}%
\bibitem [{\citenamefont {Zang}\ \emph {et~al.}(2024)\citenamefont {Zang},
  \citenamefont {Liu}, \citenamefont {Helson},\ and\ \citenamefont
  {Kumar}}]{zangStructuralConstraintsEmergence2024}%
  \BibitemOpen
  \bibfield  {author} {\bibinfo {author} {\bibfnamefont {J.}~\bibnamefont
  {Zang}}, \bibinfo {author} {\bibfnamefont {S.}~\bibnamefont {Liu}}, \bibinfo
  {author} {\bibfnamefont {P.}~\bibnamefont {Helson}},\ and\ \bibinfo {author}
  {\bibfnamefont {A.}~\bibnamefont {Kumar}},\ }\bibfield  {title} {\bibinfo
  {title} {Structural constraints on the emergence of oscillations in
  multi-population neural networks},\ }\href
  {https://doi.org/10.7554/eLife.88777} {\bibfield  {journal} {\bibinfo
  {journal} {eLife}\ }\textbf {\bibinfo {volume} {12}},\ \bibinfo {pages}
  {RP88777} (\bibinfo {year} {2024})}\BibitemShut {NoStop}%
\bibitem [{\citenamefont {Marenduzzo}\ \emph {et~al.}(2025)\citenamefont
  {Marenduzzo}, \citenamefont {Brown}, \citenamefont {Miller},\ and\
  \citenamefont {Ackland}}]{marenduzzoOscillationSIRSModel2025}%
  \BibitemOpen
  \bibfield  {author} {\bibinfo {author} {\bibfnamefont {D.}~\bibnamefont
  {Marenduzzo}}, \bibinfo {author} {\bibfnamefont {A.~T.}\ \bibnamefont
  {Brown}}, \bibinfo {author} {\bibfnamefont {C.~W.}\ \bibnamefont {Miller}},\
  and\ \bibinfo {author} {\bibfnamefont {G.~J.}\ \bibnamefont {Ackland}},\
  }\bibfield  {title} {\bibinfo {title} {Oscillation in the {{SIRS}} model},\
  }\href {https://doi.org/10.1016/j.jtbi.2025.112169} {\bibfield  {journal}
  {\bibinfo  {journal} {Journal of Theoretical Biology}\ }\textbf {\bibinfo
  {volume} {611}},\ \bibinfo {pages} {112169} (\bibinfo {year}
  {2025})}\BibitemShut {NoStop}%
\bibitem [{\citenamefont {Aliakbarian}\ and\ \citenamefont
  {{Moghimi-Araghi}}(2026)}]{aliakbarianTransitionSelforganizedCriticality2026}%
  \BibitemOpen
  \bibfield  {author} {\bibinfo {author} {\bibfnamefont {N.}~\bibnamefont
  {Aliakbarian}}\ and\ \bibinfo {author} {\bibfnamefont {S.}~\bibnamefont
  {{Moghimi-Araghi}}},\ }\bibfield  {title} {\bibinfo {title} {Transition from
  self-organized criticality towards self-organized bistability},\ }\href
  {https://doi.org/10.1016/j.physa.2025.131148} {\bibfield  {journal} {\bibinfo
   {journal} {Physica A: Statistical Mechanics and its Applications}\ }\textbf
  {\bibinfo {volume} {682}},\ \bibinfo {pages} {131148} (\bibinfo {year}
  {2026})}\BibitemShut {NoStop}%
\bibitem [{\citenamefont {Kaufman}\ and\ \citenamefont
  {Ni}(2021)}]{kaufmanQuantumScienceOpticalTweezer2021}%
  \BibitemOpen
  \bibfield  {author} {\bibinfo {author} {\bibfnamefont {A.~M.}\ \bibnamefont
  {Kaufman}}\ and\ \bibinfo {author} {\bibfnamefont {K.-K.}\ \bibnamefont
  {Ni}},\ }\bibfield  {title} {\bibinfo {title} {Quantum science with optical
  tweezer arrays of ultracold atoms and molecules},\ }\href
  {https://doi.org/10.1038/s41567-021-01357-2} {\bibfield  {journal} {\bibinfo
  {journal} {Nature Physics}\ }\textbf {\bibinfo {volume} {17}},\ \bibinfo
  {pages} {1324} (\bibinfo {year} {2021})}\BibitemShut {NoStop}%
\bibitem [{\citenamefont {Kerskens}\ and\ \citenamefont
  {L{\'o}pez~P{\'e}rez}(2022)}]{kerskensExperimentalIndicationsNonclassical2022}%
  \BibitemOpen
  \bibfield  {author} {\bibinfo {author} {\bibfnamefont {C.~M.}\ \bibnamefont
  {Kerskens}}\ and\ \bibinfo {author} {\bibfnamefont {D.}~\bibnamefont
  {L{\'o}pez~P{\'e}rez}},\ }\bibfield  {title} {\bibinfo {title} {Experimental
  indications of non-classical brain functions},\ }\href
  {https://doi.org/10.1088/2399-6528/ac94be} {\bibfield  {journal} {\bibinfo
  {journal} {Journal of Physics Communications}\ }\textbf {\bibinfo {volume}
  {6}},\ \bibinfo {pages} {105001} (\bibinfo {year} {2022})}\BibitemShut
  {NoStop}%
\bibitem [{\citenamefont {Liu}\ \emph {et~al.}(2024)\citenamefont {Liu},
  \citenamefont {Chen},\ and\ \citenamefont
  {Ao}}]{liuEntangledBiphotonGeneration2024}%
  \BibitemOpen
  \bibfield  {author} {\bibinfo {author} {\bibfnamefont {Z.}~\bibnamefont
  {Liu}}, \bibinfo {author} {\bibfnamefont {Y.-C.}\ \bibnamefont {Chen}},\ and\
  \bibinfo {author} {\bibfnamefont {P.}~\bibnamefont {Ao}},\ }\bibfield
  {title} {\bibinfo {title} {Entangled biphoton generation in the myelin
  sheath},\ }\href {https://doi.org/10.1103/PhysRevE.110.024402} {\bibfield
  {journal} {\bibinfo  {journal} {Physical Review E}\ }\textbf {\bibinfo
  {volume} {110}},\ \bibinfo {pages} {024402} (\bibinfo {year}
  {2024})}\BibitemShut {NoStop}%
\bibitem [{\citenamefont
  {Svanishvili}(2025)}]{ExploringConsciousnessPhoton2025}%
  \BibitemOpen
  \bibfield  {author} {\bibinfo {author} {\bibfnamefont {G.}~\bibnamefont
  {Svanishvili}},\ }\href {https://doi.org/10.70389/PJS.100055} {\bibinfo
  {title} {Exploring {{Consciousness}}: {{Photon Entanglement}} and {{Neural
  Communication}} - {{Premier Science}}}} (\bibinfo {year} {2025})\BibitemShut
  {NoStop}%
\bibitem [{\citenamefont {Bravo}\ \emph {et~al.}(2022)\citenamefont {Bravo},
  \citenamefont {Najafi}, \citenamefont {Gao},\ and\ \citenamefont
  {Yelin}}]{bravoQuantumReservoirComputing2022}%
  \BibitemOpen
  \bibfield  {author} {\bibinfo {author} {\bibfnamefont {R.~A.}\ \bibnamefont
  {Bravo}}, \bibinfo {author} {\bibfnamefont {K.}~\bibnamefont {Najafi}},
  \bibinfo {author} {\bibfnamefont {X.}~\bibnamefont {Gao}},\ and\ \bibinfo
  {author} {\bibfnamefont {S.~F.}\ \bibnamefont {Yelin}},\ }\bibfield  {title}
  {\bibinfo {title} {Quantum {{Reservoir Computing Using Arrays}} of {{Rydberg
  Atoms}}},\ }\href {https://doi.org/10.1103/PRXQuantum.3.030325} {\bibfield
  {journal} {\bibinfo  {journal} {PRX Quantum}\ }\textbf {\bibinfo {volume}
  {3}},\ \bibinfo {pages} {030325} (\bibinfo {year} {2022})}\BibitemShut
  {NoStop}%
\bibitem [{\citenamefont {Weber}\ \emph {et~al.}(2017)\citenamefont {Weber},
  \citenamefont {Tresp}, \citenamefont {Menke}, \citenamefont {Urvoy},
  \citenamefont {Firstenberg}, \citenamefont {B{\"u}chler},\ and\ \citenamefont
  {Hofferberth}}]{weberCalculationRydbergInteraction2017}%
  \BibitemOpen
  \bibfield  {author} {\bibinfo {author} {\bibfnamefont {S.}~\bibnamefont
  {Weber}}, \bibinfo {author} {\bibfnamefont {C.}~\bibnamefont {Tresp}},
  \bibinfo {author} {\bibfnamefont {H.}~\bibnamefont {Menke}}, \bibinfo
  {author} {\bibfnamefont {A.}~\bibnamefont {Urvoy}}, \bibinfo {author}
  {\bibfnamefont {O.}~\bibnamefont {Firstenberg}}, \bibinfo {author}
  {\bibfnamefont {H.~P.}\ \bibnamefont {B{\"u}chler}},\ and\ \bibinfo {author}
  {\bibfnamefont {S.}~\bibnamefont {Hofferberth}},\ }\bibfield  {title}
  {\bibinfo {title} {Calculation of {{Rydberg}} interaction potentials},\
  }\href {https://doi.org/10.1088/1361-6455/aa743a} {\bibfield  {journal}
  {\bibinfo  {journal} {Journal of Physics B: Atomic, Molecular and Optical
  Physics}\ }\textbf {\bibinfo {volume} {50}},\ \bibinfo {pages} {133001}
  (\bibinfo {year} {2017})}\BibitemShut {NoStop}%
\bibitem [{\citenamefont {Noh}\ and\ \citenamefont
  {Jhe}(2010)}]{nohAnalyticSolutionsOptical2010}%
  \BibitemOpen
  \bibfield  {author} {\bibinfo {author} {\bibfnamefont {H.-R.}\ \bibnamefont
  {Noh}}\ and\ \bibinfo {author} {\bibfnamefont {W.}~\bibnamefont {Jhe}},\
  }\bibfield  {title} {\bibinfo {title} {Analytic solutions of the optical
  {{Bloch}} equations},\ }\href {https://doi.org/10.1016/j.optcom.2010.01.069}
  {\bibfield  {journal} {\bibinfo  {journal} {Optics Communications}\ }\textbf
  {\bibinfo {volume} {283}},\ \bibinfo {pages} {2353} (\bibinfo {year}
  {2010})}\BibitemShut {NoStop}%
\bibitem [{\citenamefont {Mischke}\ \emph {et~al.}(2026)\citenamefont
  {Mischke}, \citenamefont {Niederprüm},\ and\ \citenamefont {Ott}}]{data}%
  \BibitemOpen
  \bibfield  {author} {\bibinfo {author} {\bibfnamefont {P.}~\bibnamefont
  {Mischke}}, \bibinfo {author} {\bibfnamefont {T.}~\bibnamefont
  {Niederprüm}},\ and\ \bibinfo {author} {\bibfnamefont {H.}~\bibnamefont
  {Ott}},\ }\href {https://doi.org/https://dx.doi.org/10.26204/data/27}
  {\bibinfo {title} {Simulating neural network criticality and resource
  dynamics with ultracold rydberg gases [dataset] (unpublished)}} (\bibinfo
  {year} {2026})\BibitemShut {NoStop}%
\end{thebibliography}
\end{document}


\title{Simulating neural network criticality and resource dynamics with Rydberg gases}
\makeatletter
\let\Title\@title
\title{Supplementary Information for\\ \Title}
\makeatother
\corauthor{Patrick Mischke\orcidlink{0000-0001-7859-8426}}{agott-publication@physik.rptu.de}
\affiliation{RPTU University Kaiserslautern-Landau, Department of Physics and State Research Center OPTIMAS, Kaiserslautern, Germany}
\affiliation{Max Planck Graduate Center with Johannes Gutenberg University Mainz (MPGC), 55128 Mainz, Germany}

\author{Herwig Ott\orcidlink{0000-0002-3155-2719}}
\affiliation{RPTU University Kaiserslautern-Landau, Department of Physics and State Research Center OPTIMAS, Kaiserslautern, Germany}

\author{Michael Fleischhauer\orcidlink{0000-0003-4059-7289}}
\affiliation{RPTU University Kaiserslautern-Landau, Department of Physics and State Research Center OPTIMAS, Kaiserslautern, Germany}

\author{Thomas Niederprüm\orcidlink{0000-0001-8336-4667}}
\affiliation{RPTU University Kaiserslautern-Landau, Department of Physics and State Research Center OPTIMAS, Kaiserslautern, Germany}

\maketitle

\section{Calculating the branching ratio}
\label{sup.sec:branching-ratio}
The goal of this section is to calculate the average number of facilitated excitations $BR$ that a seed excitation will create in the mean-field picture. For that, we assume that the seed excitation decays with rate $\gamma$ and the dynamics of a second test atom initially in the ground state and at distance $r$ from the seed atom can be described by the optical Bloch equations.

We're interested in the state of the second atom at the moment when the seed decays.
We start with the generic solution of the optical Bloch equations for an atom initially in the ground state as given in equation (10) of \cite{nohAnalyticSolutionsOptical2010}:

\begin{equation*}
    w=Ae^{-at}+\left[B\cos{(qt)}+\frac{C}{q}\sin{(qt)}\right]e^{-pt}+w_s
\end{equation*}

Where $w$ is the imbalance $w=\rho_{gg} - \rho_{ee}$. Here $w_s = \frac{\Delta^2+{\gamma^*}^2}{\Delta^2+\gamma^*(\gamma^*+\Omega^2/\gamma)}$ is the steady-state imbalance given in equation (3) of \cite{nohAnalyticSolutionsOptical2010} and $A$,$B$,$C$,$a$,$p$ and $q$ are prefactors depending on Rabi frequency $\Omega$, local detuning $\Delta(r)$, decay rate $\gamma$ and dephasing $\gamma^*$, and are explicitly given in equation (11) of \cite{nohAnalyticSolutionsOptical2010}:

\begin{align*}
    A & = \frac{1}{D}\left[-\Omega^2+(p^2+q^2)(1-w_s)\right]          \\
    B & = \frac{1}{D}\left[a(a-2p)(1-w_s)+\Omega^2\right]             \\
    C & = \frac{1}{D}\left[-\Omega^2(a-p)+a(ap-p^2+q^2)(1-w_s)\right] \\
    D & = (a-p)^2+q^2
\end{align*}

And equation (6) of \cite{nohAnalyticSolutionsOptical2010} gives the values for $a$, $p$ and $q$:

\begin{align*}
    a      & =\alpha-(S_++S_-),                     \\
    p      & =\alpha+\frac{1}{2}(S_++S_-),          \\
    q      & =\frac{\sqrt{3}}{2}(S_+-S_-)           \\
    \alpha & =\frac{2\gamma^*+\gamma}{3},           \\
    S_\pm  & =\left(R\pm\sqrt{R^2+Q^3}\right)^{1/3}
\end{align*}

Our seed atom decays at time $t$ with probability $\gamma e^{-\gamma t}$. The probability to find our test atom in the excited state once this happens can be written as $\frac{1-w}{2}$. To obtain the probability $P_\mathrm{fac}(\Omega, \gamma, \gamma*, \Delta(r))$ that the test atom gets facilitated, we integrate over all possible decay times $t$:

\begin{widetext} 
    \begin{align*}
        P_\mathrm{fac} & = \int_{0}^\infty \frac{1-w}{2}\gamma e^{-\gamma t}dt                                                                                                                                                                       \\
                       & = \frac{\gamma}{2} \int_0^\infty \left(1 - Ae^{-at} - \left[B\cos{(qt)}+\frac{C}{q}\sin{(qt)}\right]e^{-pt}-w_s\right)e^{-\gamma t} dt                                                                                      \\
                       & = \frac{\gamma}{2} \left((1-w_s)\int_0^\infty e^{-\gamma t} dt - A\int_0^\infty e^{-(\gamma + a) t} dt - B \int_0^\infty \cos{(qt)}e^{-(\gamma + p)t}dt - \frac{C}{q} \int_0^\infty \sin{(qt)}e^{-(\gamma + p)t}dt  \right) \\
                       & = \frac{\gamma}{2}\left((1-w_s)\frac{1}{\gamma} - \frac{A}{\gamma+a}-\frac{B(\gamma+p)}{(\gamma+p)^2+q^2} - \frac{Cq}{q((\gamma+p)^2+q^2))}\right)
    \end{align*}
\end{widetext}

Our derivation includes off-resonant excitation, which is always present regardless of any seed atoms. We therefore need to subtract this off-resonant rate in order to obtain the effect of our seed atom. The total branching ratio $BR$ can be derived by integrating this probability over the volume of the system and the density $n$ of ground state atoms. The local detuning $\Delta(r)$ is given by the laser detuning $\Delta$ and the $c_6$ interaction as $\Delta(r) = \Delta - \frac{C_6}{r^6}$:

\begin{widetext} 
    \begin{align*}
        BR & = \int_{R^3} n(R) \left[P_\mathrm{fac}\left(\Omega, \gamma, \gamma*, \Delta - \frac{C_6}{r^6}\right) - P_\mathrm{fac}\left(\Omega, \gamma, \gamma*, \Delta\right)\right]dV         \\
           & = 4\pi n\int_0^\infty r^2 \left[P_\mathrm{fac}\left(\Omega, \gamma, \gamma*, \Delta - \frac{C_6}{r^6}\right) - P_\mathrm{fac}\left(\Omega, \gamma, \gamma*, \Delta\right)\right]dr
    \end{align*}

    Setting $BR=1$ allows solving for the critical density $n_c$:

    \begin{equation}
        n_c = \frac{1}{4\pi \int_0^\infty r^2 \left[P_\mathrm{fac}\left(\Omega, \gamma, \gamma*, \Delta - \frac{C_6}{r^6}\right) - P_\mathrm{fac}\left(\Omega, \gamma, \gamma*, \Delta\right)\right]dr}
        \label{sup.eq:critical-density}
    \end{equation}

    A numerical solution of this integral is shown as yellow line in Fig.\,2.
\end{widetext}

\section{Interaction potential}

We calculate the Rydberg-Rydberg interaction using the pairinteraction \cite{weberCalculationRydbergInteraction2017} software package. The potential curves are shown in Supplementary Fig.\,\ref{sup.fig:interaction-potential}. Fitting a $C_6/r^6$ potential to the resulting curve allows us to obtain a value of $C_6 = \SI{0.111}{\giga\hertz\micro\meter^6}$ and a facilitation distance of $r_\mathrm{fac} =\SI{1.185}{\micro\meter}$ at a laser detuning of $\Delta = \SI{40}{\mega\hertz}$.

\begin{figure}
    \includegraphics{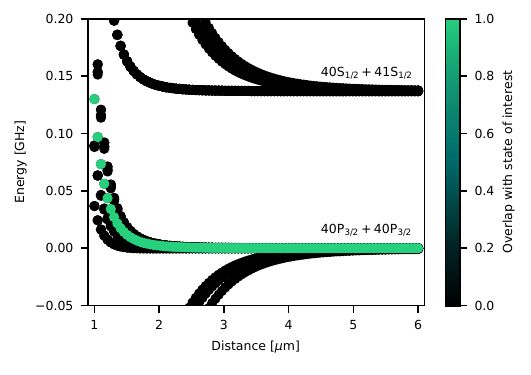}
    \caption{Rydberg interaction potential. Calculation is done using the pairinteraction software package \cite{weberCalculationRydbergInteraction2017}. The different $40\mathrm{P}$ curves correspondent to different $m_j$ quantum numbers. Which one our coupling laser excites depends on the angle between inter-atomic axis (which defines the quantization axis, as the Rydberg-Rydberg interaction is our largest energy scale) and laser polarization. The $40\mathrm{S} + 41\mathrm{S}$ pair state at $+\SI{137}{\mega\hertz}$ detuning causes additional features in the detuning dependent phase diagram shown in Fig.\,2.}
    \label{sup.fig:interaction-potential}
\end{figure}

\bibliography{bibliography}